\documentclass{aa} % for a standard version
\usepackage{graphicx}
\usepackage{txfonts}
\usepackage{natbib}
\usepackage[citecolor=blue,colorlinks=true,linkcolor=blue,urlcolor=blue]{hyperref}
\graphicspath{{./}{figures/}}

\begin{document}

   \title{SRG/eROSITA and XMM-Newton observations of Vela Jr}

   %\subtitle{I. Overviewing the $\kappa$-mechanism}

\author{Francesco Camilloni
    \inst{1}\thanks{\email{fcam@mpe.mpg.de}}
    \and Werner Becker \inst{1}\inst{,2}
    \and Peter Predehl \inst{1}
    \and Konrad Dennerl \inst{1}
    \and Michael Freyberg  \inst{1}
    \and Martin G. F. Mayer  \inst{1}
    \and Manami Sasaki \inst{3}
}

\institute{Max-Planck-Institut für extraterrestrische Physik, Giessenbachstraße, 85748 Garching, Germany
 \and
Max-Planck-Institut für Radioastronomie, Auf dem Hügel 69, 53121 Bonn, Germany
 \and
 Dr. Karl Remeis Observatory, Erlangen Centre for Astroparticle Physics, Friedrich-Alexander-Universität Erlangen-Nürnberg,
Sternwartstrasse 7, 96049 Bamberg, Germany
}
   \date{Received 16 November 2022 / Accepted 13 March 2023}

% \abstract{}{}{}{}{}
% 5 {} token are mandatory

  \abstract{The Vela supernova remnant complex is a region containing at least three  supernova remnants: Vela, Puppis A, and Vela Jr. With the launch of the spectro-imaging X-ray telescope eROSITA on board the  Spectrum Roentgen Gamma (\textit{SRG}) mission, it became possible to observe the one degree wide Vela Jr in its   entirety. Although several previous pointed \textit{Chandra} and \textit{XMM-Newton} observations are available, it is only the second time after the \textit{ROSAT} all-sky survey that the whole remnant was observed in X-rays with homogeneous sensitivity.}
  {Vela Jr is  one of the few remnants emitting in the TeV band, making it an important object in shock acceleration studies. However, the age and distance determination using X-ray emission is largely hampered by the presence of the Vela SNR along the same line. With the  eROSITA data set our  aim is to characterize the emission of Vela Jr and distinguish it from Vela emission, and also to characterize  the spectral emission of the inner remnant.}
  {We processed the eROSITA data dividing the whole remnant into seven different regions. In addition, images of the whole remnant were employed to pinpoint the position of the geometric center and constrain the proper motion of the CCO. We also employed archival \textit{XMM-Newton} pointed observations of the NW rim to determine the cutoff energy of the electrons and the expansion velocity.}
  {We find the magnetic field can vary between 2 $\mu$G and 16 $\mu$G in the NW rim. We also find that the remnant spectrum is uniformly featureless in most regions, except for two inner regions where an extra thermal model component improves the fit. We obtain new coordinates for the geometric remnant center, resulting in a separation of only 35.2 $\pm$ 15.8" from the position of the CCO. As a result, we reinforce the association between the CCO and a proposed faint optical--IR counterpart.}
  {}

   \maketitle
%
%-------------------------------------------------------------------

\section{Introduction \label{sec:intro}}
The increase in angular resolution and sensitivity of $\gamma$-ray detectors like the High Energy Stereoscopic System (HESS) \citep{2018A&A...612A...6H} produced a considerable step forward in multiwavelength studies of supernova remnants (SNRs) in the last decade. The main mechanism responsible for particle acceleration in SNRs is diffusive shock acceleration \citep[cf.][]{1987PhR...154....1B}. This is particularly true for young remnants, which are often characterized by a featureless continuum spectrum from the radio to the X-ray and $\gamma$-ray bands \citep[e.g.,][]{Bamba05a,2016A&A...593A..26M,2018PASJ...70...77O}, indicating that the detected radiation is  nonthermal synchrotron emission. Moreover, multiband fits like those presented by \cite{2010ApJ...721.1492P} and \cite{Kishishita2013} have allowed us to put constraints on the underlying electron distribution, relevant to characterizing the Galactic cosmic ray population. By the middle of 2022, 
the online catalog of TeV sources at the University of Chicago \citep{2008ICRC....3.1341W}
lists 95 entries.\footnote{\url{http://tevcat.uchicago.edu}} Among them, only eight are supernova remnants, of which Vela Jr (also known as RX J0852.0-4622 or G266.2-1.2) is  the one with the largest extent. All these remnants are shell-type \citep{2018A&A...612A...1H,2018A&A...612A...3H}, excluding contributions of pulsars or pulsar wind nebulae (PWN). 
Vela Jr was discovered by \cite{1998Natur.396..141A} using data from the \textit{ROSAT} all-sky survey, which was the first all-sky survey performed by an imaging X-ray telescope. With the launch of the Spectrum-Röntgen-Gamma (\textit{SRG}) mission in June 2019 \citep{2021A&A...656A.132S}, for which the    extended ROentgen Survey with an Imaging Telescope Array (eROSITA)  is the soft X-ray instrument on board \citep{2021A&A...647A...1P}, another possibility for observing the two-degree wide Vela Jr remnant in its entirety by a spectro-imaging X-ray telescope became available. In this work we report on the results of an analysis of this data. In order to increase the photon statistics  for our analysis, we also made use of archival \textit{XMM-Newton} calibration data taken on the  northwestern (NW) region of the remnant between 2001 and 2021. 

Soon after the discovery of Vela Jr, \cite{1999A&A...350..997A} found evidence in favor of a remnant distance of less than 1 kpc and an estimated remnant age of 680 yr. This study relied mostly on data from the \textit{COMPTEL} $\gamma$-ray observatory and an apparent $^{44}$Ti 
signal from Vela Jr \citep{1998Natur.396..142I,1999ApJ...514L.103C}. The isotope  $^{44}$Ti has a half life of only 60 yr, so that its presence lends strong support for a young remnant age. However, the significance of that $^{44}$Ti emission line was at the 3$\sigma$ level only, \citep{2006NewAR..50..540R} and later observations indeed did not confirm its existence  \citep{2009PASJ...61..275H,2020A&A...638A..83W}. Today, more than 30 years after the discovery of Vela Jr in the \textit{ROSAT} all-sky survey, there is still no agreement about its distance and age. One of the main observational problems with this remnant is its location along the line of sight of the much more extended Vela SNR. The presence of Vela SNR makes the spectral analysis of Vela Jr very complex. Especially in the soft X-ray band, spectral modeling is limited by the problem of disentangling the emission from the two remnants. Most of the spectral studies in X-rays conducted so far have ignored the energy range below 1 keV for that specific reason \citep[cf.][]{2005A&A...429..225I, 2008ApJ...678L..35K, 2010ApJ...721.1492P,2016PASJ...68S..10T}. 

Using \textit{ASCA} data, \cite{2001ApJ...548..814S} estimated a remnant distance 
of 1-2 kpc based on the significantly higher column density measured for Vela Jr than for the Vela SNR. If correct, Vela Jr would be a background object relative to the Vela SNR, which a parallax measurement of the central pulsar placed at a distance of $287^{+19}_{-17}$ pc \citep{2003ApJ...596.1137D}. Conversely, from the estimate given in \cite{1999A&A...350..997A}, it is possible that Vela Jr is a foreground object. Several later studies concluded that Vela Jr was much closer than 1-2 kpc. For example, \cite{Bamba05b} studied the filamentary structure of the NW rim with \textit{Chandra} and obtained a distance of around 0.33 kpc, which is almost compatible with the distance of the Vela SNR. \cite{2008ApJ...678L..35K} followed a different approach by studying the expansion rate of the NW rim in order to test the scenarios presented by \cite{1999A&A...350..997A} and \cite{2001ApJ...548..814S}: the rather low expansion rate of $0.84 \pm 0.23$" yr$^{-1}$ made them conclude that Vela Jr was most likely a supernova remnant of age $1700-4300$ years,  located at a distance of around 750 pc. However, also these authors were not able to completely rule out the hypothesis that Vela Jr  is indeed a very young remnant that was rapidly decelerated, for example in the interaction with a dense interstellar medium.

At X-ray energies, the northwestern region of Vela Jr was found to show a filamentary trailing emission behind a shock front, tentatively associated with a reverse shock \citep{2005A&A...429..225I}. This region also appears to be the brightest part of the remnant. To explore it in more detail, it was in the focus of several \textit{XMM-Newton} and \textit{Chandra} observations. \textit{XMM-Newton} in particular   monitored the NW region in regular observations between 2001 and 2021.
%
% it brings the focus's away (to my opinion) if we include the following %
%: together with the almost spherical shape, these could be two factors indicating a young age. Indeed, for an old %remnant is very likely to expect deviation from spherical symmetry due %to interaction with the ISM. Moreover, reflection features created by %the reverse shock are expected to be dissipated if the remnant is old %\citep{2012A&ARv..20...49V}. 
%
%
\cite{2015ApJ...798...82A} measured the expansion rate of the NW rim, comparing two \textit{Chandra} observation performed in  2003 and 2008. They obtained an expansion rate of $0.42 \pm 0.10$ arcsec yr$^{-1}$, which is half that measured using \textit{XMM-Newton} by \cite{2008ApJ...678L..35K}. 
Inserting this expansion rate as a constraint on a numerical 
simulation based on the models of \cite{1999ApJS..120..299T}, 
%along with the initial kinetic energy ($10^{49}-10^{51}$ erg), ejecta mass ($10^{0}-10^{2}$ M$_{\odot}$), mass density distribution of the ejecta (m=$0.50-0.79$), mass density distribution of the ambient medium (n$_{\text{0}}$:$10^{-5}-10^{0}$ cm$^{-3}$) and evolutionary state (t/t$_{\text{ch}}$: 0.01, 0.02,..., 9.99, where t$_{\text{ch}}$ =6.18(E$_{0}/10^{51}$ erg)$^{-1/2}$(M$_{\text{ej}}$/10 M$_{\odot}$)$^{5/6}$(n$_{\text{0}}$/0.1  cm$^{-3}$)$^{-1/3}$),
the authors concluded that Vela Jr has an age in the range $2.4 - 5.1$ kyr. Using synchrotron cutoff frequency measurements from \cite{2010ApJ...721.1492P}, these authors set a lower limit of 0.5 kpc to the distance of Vela Jr.

As described above, there is no clear tendency 
for the distance estimate and age of Vela Jr. Summarizing the results from the literature sets the lower and upper limits for the remnant distance at about $0.3$ and $2$ kpc. While there is general consent that the remnant is young, whether it is a historical remnant of age $\sim 700$ years or whether its age is more comparable with that of Puppis A \citep[around 4000 years; see][]{Winkler88,Mayer20} is still under debate.

In this work, we use  eROSITA all-sky survey data to conduct a detailed spectro-imaging analysis of the X-ray emission of Vela Jr. Our data set constitutes  the most sensitive observation covering the entire remnant in the $0.2-10.0$ keV energy band, with an angular resolution of 26 arcsec in survey mode and 15 arcsec in pointed mode (NW rim). It further offers a relatively uniform exposure over the extent of Vela Jr, eliminating the need to create mosaics from many individual observations. In addition,
to better constrain the emission from the NW region of Vela Jr, we included archival \textit{XMM-Newton} data in our analysis. This data were taken for calibration purposes\footnote{All the XMM observations except the one performed in 2001  were taken in calibration mode (both PN and MOS). In this configuration mode, the decaying radioactive source of calibration is illuminating the detector. This can potentially disturb spectral measurements if no special care is taken for it during the data analysis.} on a yearly basis between 2001 and 2021. 

Our paper is organized as follows.
Section \ref{sec:Data} presents the basic characteristics of our data sets and describes initial data reduction steps taken to ensure its correct treatment. We describe our methods and results in imaging and spectroscopic analysis in Sect. \ref{sec:Spatial_Analysis} and Sect. \ref{sec:Spectroscopy};  the core results of spatially resolved spectroscopy of Vela Jr are presented in Sect. \ref{sec:Spectroscopy}. In Sect. \ref{sec:discussion_distance} we present our distance analysis. Finally, in Sect. \ref{sec:Results_Discussion} we summarize our results and discuss their physical implications. 

\section{Data reduction \label{sec:Data}}

\begin{table*}[]
    \caption{Observations analyzed in the paper. PN, MOS1 and MOS2 are the detectors mounted on \textit{XMM-Newton}. eRASS:4 denotes the abbreviation of the stacked eROSITA all-sky survey data for the surveys $1-4$.
    }
    \label{tab:observations}
    \centering
    \begin{minipage}{\textwidth}
  \centering
\begin{tabular}{lllllllll}%
Year & ObsID& Instrument & Observation Mode & Exposure (ks) & Pointing \\
\hline\\[-0.9ex]%
2001 & 0112870301 & PN, MOS1, MOS2 & Medium Filter &  31.9 & NW rim\\[1ex]%
2009 & 0412990601 & PN, MOS1, MOS2 & CalThin, CalThin, CalThin &  127 & NW rim\\[1ex]%
2010 & 0412990701 & PN, MOS1, MOS2 & CalThin, CalMedium, CalMedium & 66.5 & NW rim \\[1ex]%
2019 & 0810890201 & PN, MOS1, MOS2 & CalThin, CalMedium, CalMedium & 62.9 & NW rim\\[1ex]%
2020 & 0810890301 & PN, MOS1, MOS2 & CalThin, CalMedium, CalMedium & 62.9& NW rim \\[1ex]%
2021 & 0810890501 &  PN, MOS1, MOS2 & CalThin, CalMedium, CalMedium & 63.4 & NW rim\\[1ex]%
\hline\\[-0.9ex]%
2019 & 700039 & eROSITA& CalPV& 60.0 & NW rim \\[1ex]%
\hline\\[-0.9ex]%
2019-2021 & eRASS:4 & eROSITA& Survey&  1.27\footnote{Total not-vignetting corrected exposure on-source time} & -  \\[1ex]%
% 2019-2021 & eRASS:4 & eROSITA& Survey&  1.27\tablefootmark{a} & -  \\[1ex]%
\end{tabular}%
% \tablefoot{\tablefoottext{a}{Total not-vignetting corrected exposure on-source time}}
\end{minipage}
\end{table*}

\subsection{eROSITA data reduction\label{sec:eROSITA_reduction}}
Vela Jr was in the eROSITA field of view during four consecutive all-sky surveys performed between December 2019 and December 2021. The unvignetted exposure time of all survey observations covering Vela Jr sum up to about 1270 sec. The vignetting-corrected exposure averaged over the size of the remnant for the full $0.2-8.0$ keV band is 658 sec. In addition to the survey observations, eROSITA observed the NW rim of Vela Jr during its CalPV phase for an exposure time of 60 ksec. This observation was focused on the same position as the \textit{XMM-Newton} calibration observations taken between 2001 and 2019. This CalPV eROSITA observation was originally scheduled for the purpose of inter-calibrating its detector response with that of \textit{XMM-Newton}, and was performed on October $29-30$, 2019. A procedure very similar to that described in Section \ref{sec:XMM_reduction} was followed for the data reduction of the eROSITA survey and CalPV data. We used the eROSITA science analysis software (eSASS) version 211214\footnote{The software version is denoted according to the date when it was released, in this case 14.12.2021} \citep{2022A&A...661A...1B} with the calibration files CALDB 2021Q4 for our data analysis. 
%
%We employed the CalPV data to produce the images shown in Figure %\ref{fig:eROSITA_CalPV_image_a} and Figure %\ref{fig:eROSITA_CalPV_image_b}. 

eROSITA consists of seven telescope modules (TMs), each equipped with its own detector in the focal plane. A detailed description of the instrument is given in \cite{2021A&A...647A...1P}. Due to a light leak from the backside of the focal plane, TM5 and TM7 can be less suitable for spectroscopic analysis for certain Sun angle constraints, making the calibration of these modules less solid (at that time of the mission). This problem  primarily affects the low energy sensitivity \citep{2021A&A...647A...1P}. For the spectral analysis of the Vela Jr data we therefore employed only the merged data from   TMs 1, 2, 3, 4, and 6. The cleaned photon event files were extracted using the eSASS command  \texttt{evtool} to filter the good time intervals, and selecting all the available detection patterns (PATTERN=15). We then extracted spectra, background, redistribution matrix file (RMF), and ancillary response file (ARF) using the \texttt{srctool} tasks. 

\subsection{XMM-Newton data reduction\label{sec:XMM_reduction}}

The \textit{XMM-Newton} data used in our analysis of the NW region are
summarized in Table \ref{tab:observations}. These data were reduced using the XMM Science Analysis System (SAS version 19.1.0) together with the latest calibration files. We first created filtered event files. To do so, we generated one light curve per observation to look for the presence of solar flaring events, and cleaned the data by accepting only the good time intervals when the sky background was low. The same process was applied for both EPIC-PN and EPIC-MOS data, whose spectra were employed in the final spectral analysis, which we restricted to the $0.3-10$ keV energy band. After this cleaning procedure, we extracted the images and used them to  spatially divide the NW shock region into seven sectors. The regions were tailored in accordance with the regions selected in the work by \cite{2010ApJ...721.1492P}. We finally produced spectra for each of the seven sectors with the highest possible statistics , selecting events with PATTERN<=4,12 for PN and MOS and FLAG==0 settings in the filtering expression option of the \texttt{evselect} command. For each region a source spectrum, background, RMF and ARF were extracted using the commands \texttt{evselect}, \texttt{rmfgen}, and \texttt{arfgen}, respectively. Since we modeled the background (see Appendix \ref{app:XMM}), we did not apply any grouping to the spectra. The same procedure was applied for all observations listed in Table \ref{tab:observations}, which sum up to a total exposure time of   more than 410 ksec. 

\section{Spatial analysis \label{sec:Spatial_Analysis}}
\begin{figure}
    \centering
    \includegraphics[scale=0.5]{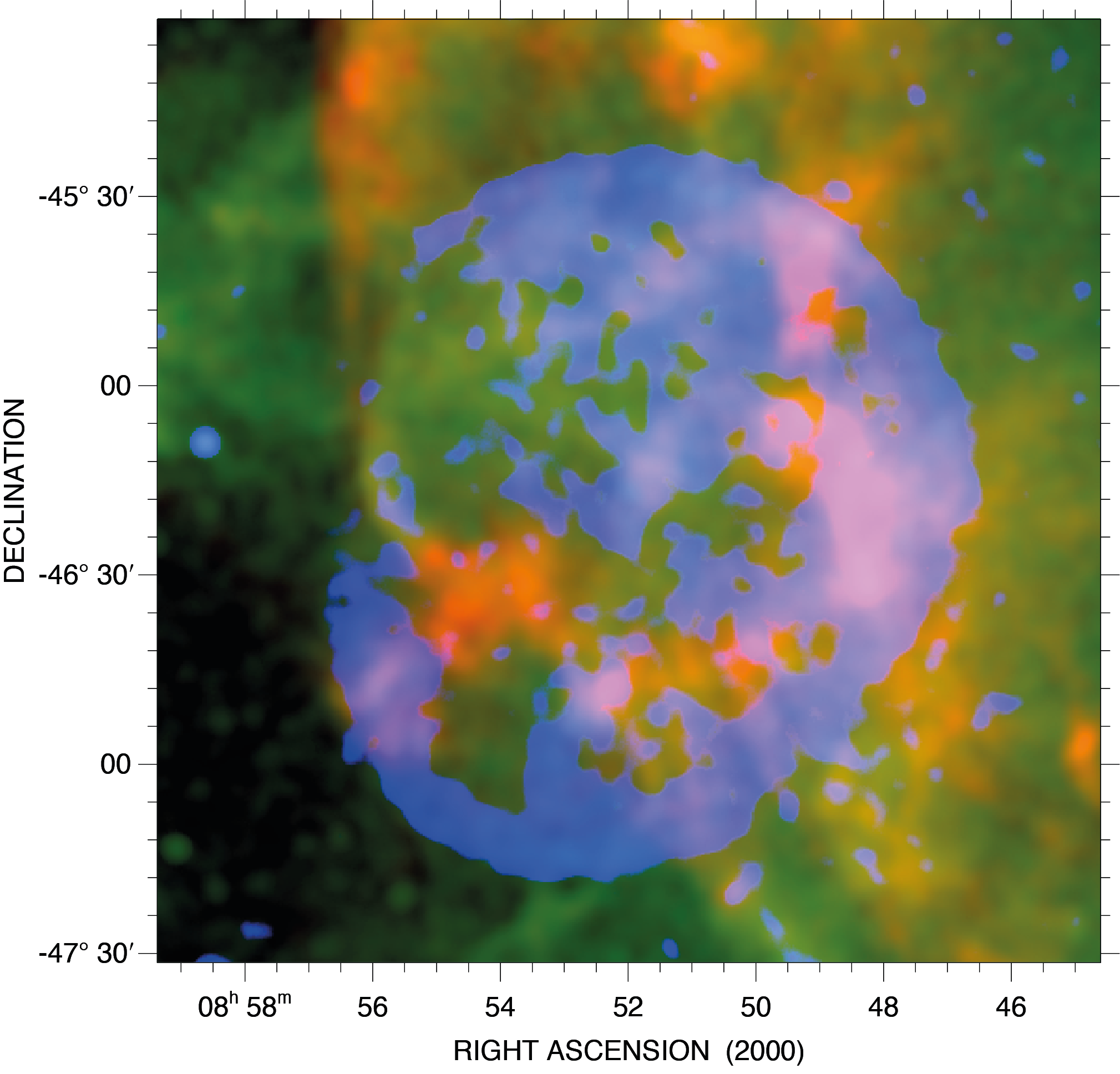}
    \caption{False color image of Vela Jr. The energy of the X-ray photons are color-coded: red ($0.2-0.7$ keV), green ($0.7-1.2$ keV), and blue ($1.2 - 8.0$ keV). In order to enhance the visibility of the diffuse X-ray emission,  the images of the different bands were convolved with a Gaussian filter (expressed in pixels; red: radius = 7 , $\sigma = 3.5$; green: radius = 8 , $\sigma= 4$; blue: radius=8 , $\sigma=4$). 
    \label{fig:VelaJr_RGB}}
\end{figure}

In survey mode eROSITA provides an  almost unlimited field of view, and
thus allowed us to map the whole Vela Jr remnant with an \textit{XMM-Newton}-like sensitivity. Figure \ref{fig:VelaJr_RGB} depicts a  multispectral image of Vela Jr as obtained from stacked eROSITA data of all four sky surveys (eRASS:4) taken between December 2019 and December 2021. Impressively, the RGB color-coding reveals for the first time the different emission contributions of the Vela SNR (in red and green) and of Vela Jr (in blue) over its entire extent. 

%Figure1

\begin{figure}
    \centering
    \includegraphics[scale=0.5]{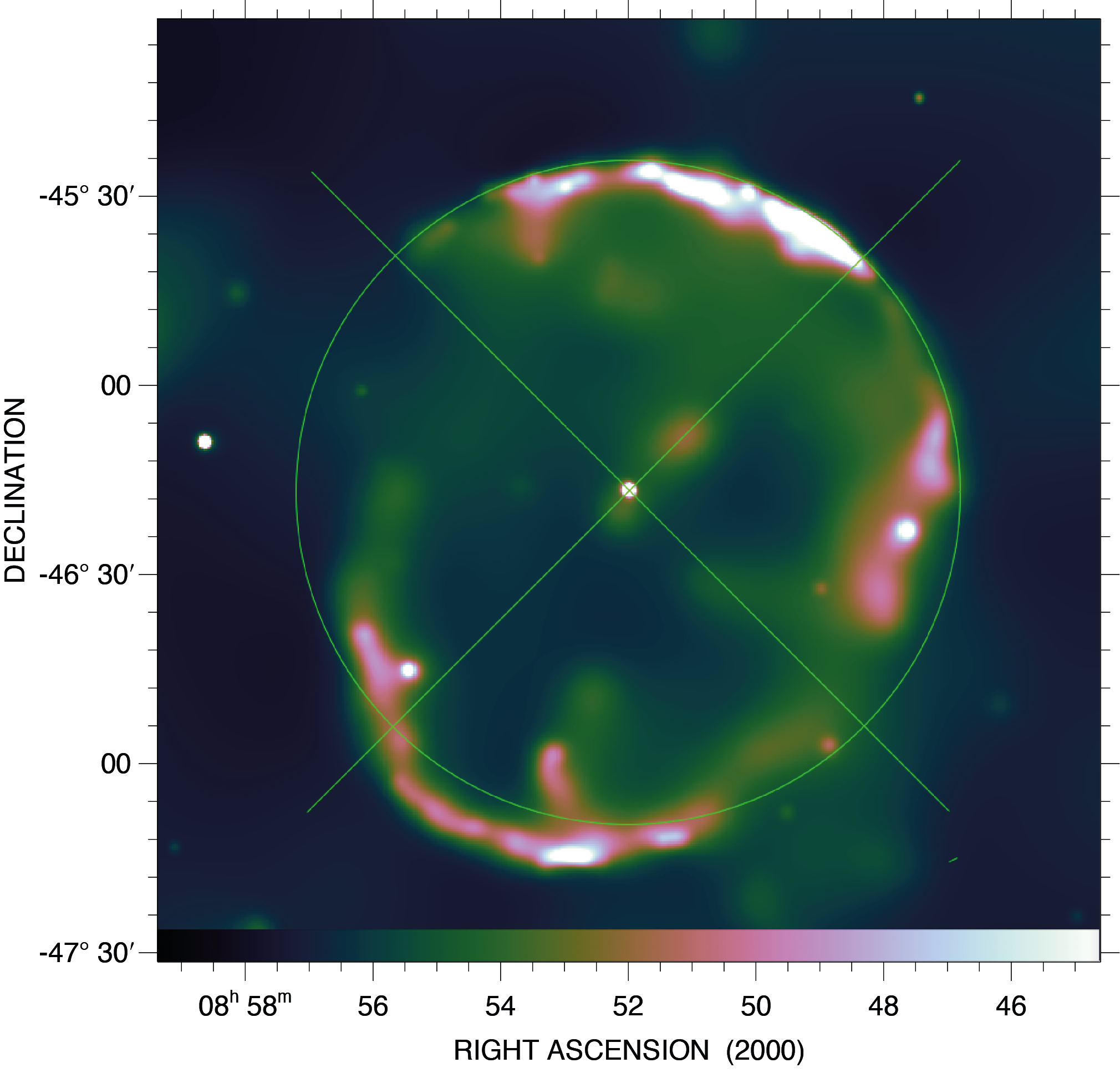}
    \caption{Intensity representation of the emission of Vela Jr and its CCO J085201.4-461753 in the $1.1-8$ keV band as seen by eROSITA's TM1--TM7.  The colors are linearly distributed so that black corresponds to a pixel intensity value of 0 and white to 3 cts/pixel. The point source at the southeastern rim of the Vela Jr remnant is the 65 ms pulsar PSR J0855-4644. %\citep{2013A&A...551A...7A}.
    \label{fig:geometric_center}}
\end{figure}

\begin{figure*}
    \centering
    \includegraphics[scale=0.9]{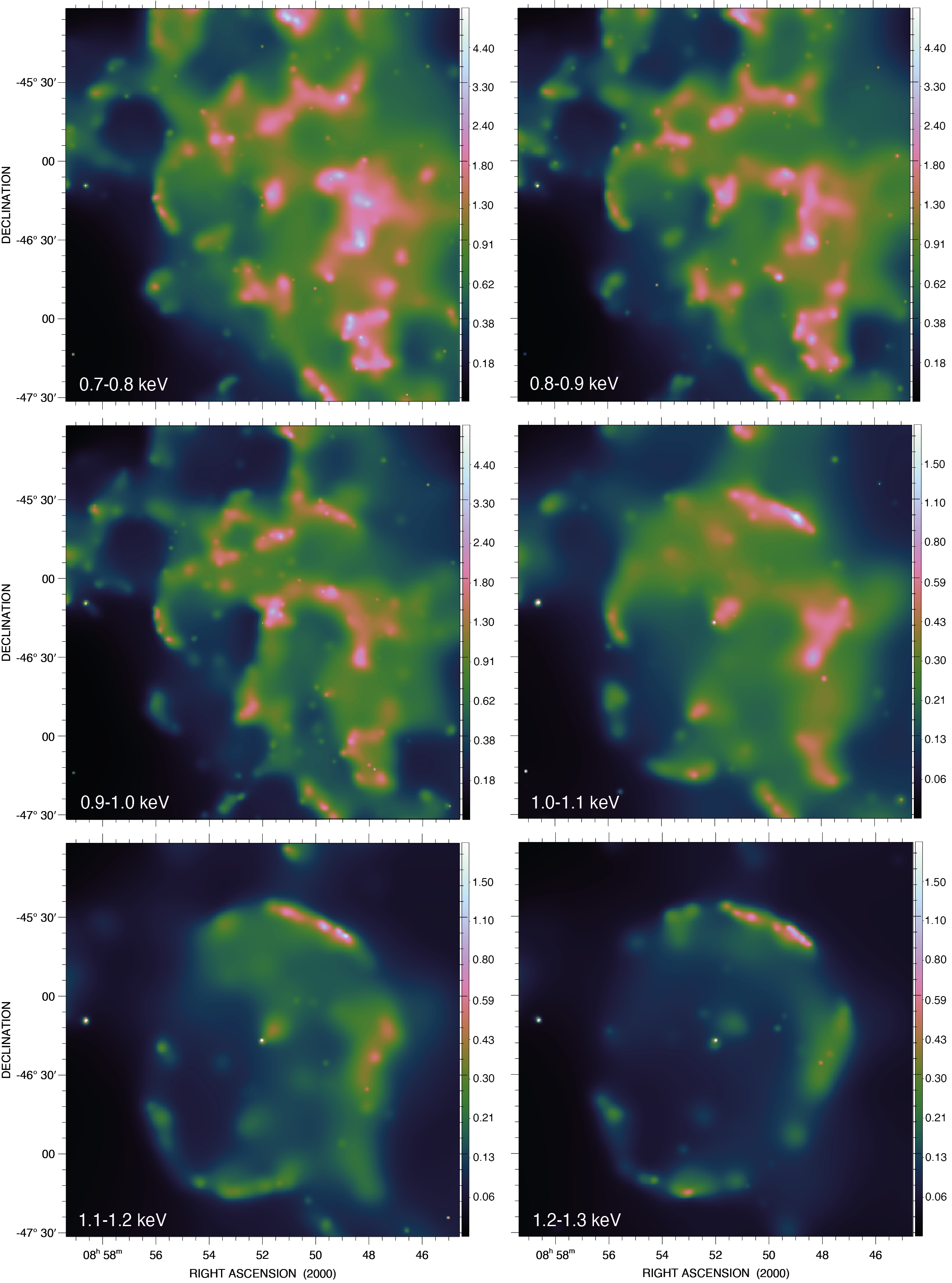}
    \caption{Energy-selected eROSITA (TM1-TM7) view of the Vela Jr region. The images are centered on the position of the CCO. Intensity vs.\ color are distributed according to the ASINH-function. The color vs. cts/pixel distribution in each image is depicted. For the $0.7-1.0$ keV images a scale of $0-6$ cts/pixel, and for the $1.1-1.3$ keV images a scale of $0-2$ cts/pixel were used. Below $\sim 1$ keV emission from the Vela SNR dominates. Above $\sim 1$ keV the nonthermal emission from Vela Jr starts to stand out above the emission from Vela SNR. 
    \label{fig:VelaJr_sliced}}
\end{figure*}

Having data of that high quality of the entire remnant allowed us for the first time after the \textit{ROSAT} all-sky survey to address the question of the remnant's geometrical center and to compare it with the position of the central compact object (CCO) CXOU J085201.4-461753 \citep[cf.][and references therein]{2006astro.ph..7081B}. The first measurement of the geometrical center of Vela Jr was obtained by \cite{1998Natur.396..141A}. Our measurement is based on the eRASS:4 intensity image shown in Figure \ref{fig:geometric_center}. 
%
%Figure2
The image was created from all events recorded within the $1.1-8.0$ keV energy band using all seven telescope modules. In this energy band Vela Jr stands out most clearly above the Vela SNR emission. To demonstrate this, we applied an energy-sliced image analysis within the $0.7-1.3$ keV band, comparing the emission contributions of the Vela SNR and Vela Jr in steps of 0.1 keV. The result is depicted in Figure  \ref{fig:VelaJr_sliced}. In order to enhance the visibility of diffuse emission in these images while leaving point sources unsmoothed to the greatest possible extent, we applied the adaptive smoothing algorithm of \cite{2006MNRAS.368...65E} with a Gaussian kernel allowed to vary between $\sigma_{\text{min}}=3.0$ pixels and $\sigma_{\text{max}}=4.0$ pixels. As can be seen in Figure \ref{fig:geometric_center}, the north to northeastern part of the remnant appears almost perfectly circular. Assuming that only this part of the remnant tracks the position of the geometric remnant center,
we fitted the upper half only with a circle so that it follows closely the outer remnant boundary. In the first instance we did this fitting by eye. To overcome possible subject bias,  these fits were performed independently by several members of the team. In Section \ref{sec:Results_Discussion} we discuss the choice of using  this part of the remnant alone to determine the geometric center. From all measurements, we computed the mean and the $1 \sigma$ standard deviation and found the center of the circle at RA=$08$h $51$m $58$s($\pm$1s), DEC= -$46$$^{\circ}$$17$'$52$''($\pm 22$''). The radius of the fitted circle is 0.95$^{\circ}\pm 0.02^{\circ}$.  The best-fitting circle is indicated in Figure \ref{fig:geometric_center}. The two lines indicated in Figure \ref{fig:geometric_center} connect the four corners of the square exactly surrounding that best-fitting circle. As can be seen, the lines cross exactly at the location of the CCO. In eROSITA we find the position of the CCO at RA=$08$h $52$m $01.42$s, DEC= $- 46^{\circ}$ $17$' $54.52$", which is 1.26" different from the position RA=$08$h $52$m $01.4$s, DEC= $- 46^{\circ}$ $17$' $53.3$" obtained with a typical error of 0.6" in a \textit{Chandra} observation by \cite{2001ApJ...559L.131P}. Based on this CCO position, the separation between the center of the fitted circle and the CCO is only 35.2 $\pm$ 15.8", which is within the $2 \sigma$ uncertainty range in agreement with zero proper motion. This result was supported by applying an alternative technique based on the edge detection filter $EDGE\_DOG$\footnote{\url{https://www.l3harrisgeospatial.com/docs/edge\_dog.html}} to Fig. \ref{fig:geometric_center} by using the Interactive Data Analysis Language IDL. This IDL function determined the outer boundary of Vela Jr on which we overlaid a circle region in ds9. The radius and geometrical center of this circle  turned out to be in full agreement with the results described above. Although the very southern part of the remnant is not well described by the circle that fits the northern part of the remnant, the fact that the position of the CCO is in agreement with the center of the best-fitting circle makes it appealing to speculate that it is still at its birthplace and has not not moved significantly from there since its formation. 
%
%Figure3

 \cite{2019MNRAS.486.5716M} measured an upper limit of 10 mas/yr for the proper motion of an infrared candidate counterpart for the CCO in the center of Vela Jr. Employing the geometric center value given by \textit{ROSAT} and the CCO position obtained with \textit{Chandra} \citep{2019MNRAS.486.5716M}, the separation is close to 4 arcmin, while, conversely, it should be comprised between 24" and 51" with an age of $2.4-5.1$ kyr  and proper motion of 10 mas/yr. From our measurements, if only the NW rim is considered when fitting a circle to the rim, we find that the geometric center of Vela Jr is very close to the CCO position, with a separation of (35.2  $\pm$ 15.8)", so this is in accordance with the proper motion limit set by the infrared counterpart. Using the transverse velocity relation employed in \cite{2019MNRAS.486.5716M}
\begin{equation}
V_{T,100} \sim 5 \times \mu_{1} \times d_{100}    
\label{eq:proper_motion}
,\end{equation}
where d$_{100}$ is the distance in units of 100 pc,  V$_{T,100}$ the velocity in units of 100 km/s and $\mu_{1}$ the proper motion in units of 1 arcsec/yr, it is possible to infer information on the velocity of the CCO.
The (35.2  $\pm$ 15.8)" of separation measured in this paper implies an upper limit on the proper motion of (14.7 $\pm$ 6.6) mas/yr, if an age of 2400 yr is assumed for the remnant. Assuming a distance of 750 pc \citep{2008ApJ...678L..35K}, we infer a transverse velocity v$=( 55\pm 25)$ km/s. If instead an age of 5100 years is assumed  (upper limit on the age range given by \citealt{2015ApJ...798...82A}), the proper motion is around (6.9 $\pm$ 3.0) mas/yr, which translates into a velocity of v$=(26 \pm 12)$ km/s at the same distance. 
In both cases the value is fully in accordance with the upper limit on the proper motion set by \cite{2019MNRAS.486.5716M} for the possible IR counterpart. Nevertheless, we note the number of uncertainties present in our analysis: the broad range of values for the age of the remnant, the uncertainty in our determination of the geometric center, and, for the velocity, the distance. 

\section{Spectral analysis \label{sec:Spectroscopy}}

All the spectra were fitted with the software PyXSPEC, the Python interface of XSPEC \citep{1996ASPC..101...17A}. The given errors correspond to the 1$\sigma$ confidence interval and the statistic employed is CSTAT \citep{Cash1979}. Solar abundances were set to those of \cite{2000ApJ...542..914W}.

%Given the considerable complexity of the background modeling, we put in Appendix the detailed description of the procedure. After the background modeling, for both the analysis we followed a Monte Carlo Markov Chain (MCMC) based approach, employing the Bayesian analysis Python library \texttt{emcee} \citep{2013PASP..125..306F}, in order to obtain reliable constraints  on the parameters. Indeed several papers in the last few years demonstrated the need of more advanced statistical tools to study X-ray spectra (see for example \cite{2001ApJ...548..224V} for a general discussion about Bayesian Analysis approach in X-ray astronomy or \cite{2010ApJ...724L.161B} for an application to SNR studies using X-ray): the main advantage over the traditional fitting technique is the capability to probe more in depth the space of parameters. This translates into a more robust determination of the parameters and the errors associated to them. A fundamental property of the Bayesian approach is to highlight immediately the degeneracy between the model parameters. The need of this choice was the spatial variability of the background, as clearly shown in Figure \ref{fig:VelaJr_RGB}, and the very high degeneracy between the emission of Vela and Vela Jr below 1 keV (Figure \ref{fig:VelaJr_sliced}). We initialised our walkers with Gaussian distribution centered on best fit parameters and we employed logarithmically uniform priors on the model components left free to vary during the run. More details on the fitting procedure are given in Appendix.

\subsection{The eROSITA survey data eRASS:4 of Vela Jr\label{sec:eRASS4_dataset}}
In this chapter we present a systematic spectral study of Vela Jr by using the stacked data of the first four eROSITA all-sky surveys (eRASS:4). 

\subsubsection{Voronoi binning approach \label{sec:Voronoi}}

We gained a first impression of the global characteristics of nonthermal emission across Vela Jr via a Voronoi binning approach. This analysis was carried out %with the main goal of briefly characterizing thermal and non-thermal emission in the Vela Jr remnant,
%in the overlapping Vela SNR (M. Mayer et al., in prep.),  but 
in order to provide a useful look at the nonthermal shell of Vela Jr, which is agnostic to any morphological features in the emission. 
We followed a similar approach as in \citet{Mayer22}, which can be summarized as follows. We ran an adaptive Voronoi tessellation algorithm \citep{Vorbin}, with a target $S/N = 100$ on the eRASS:4 count image of Vela Jr, which was extracted in the $0.2-2.3\,\mathrm{keV}$ band. The resulting Voronoi bins were fed into {\tt srctool} to extract spectra from each region using only data from telescope modules which have an on-chip filter (i.e., TMs 1, 2, 3, 4, 6). These spectra were then fitted in the energy range $0.2-8.5\,\mathrm{keV}$ with a source model combining foreground absorption \citep{2000ApJ...542..914W} with a thermal contribution from shocked plasma with non-equilibrium ionization \citep{2001ApJ...548..820B} to fit the contribution from the Vela SNR. This was combined with a power law component that accounts for the presence of nonthermal emission from the Vela Jr remnant itself. In order to avoid the challenging task of independently constraining the absorption toward Vela Jr (see Sect.~\ref{sec:NW_only}), we fixed the absorption of this nonthermal component to that of the CCO, $N_{\rm H} = 3.5\times10^{21}\,\mathrm{cm^{-2}}$ \citep{2006astro.ph..7081B}.
In Xspec, the combined source model is thus expressed as \texttt{TBabs*vpshock+TBabs*powerlaw}.
The background contribution was  modeled using a combination of three templates, reflecting the instrumental background \citep{Yeung2023}, the extragalactic X-ray background due to unresolved AGN \citep{2004A&A...419..837D}, and a combination of thermal emission models. The last consists of components taking into account emission from the local hot bubble, the Galactic halo, and unresolved emission from the Galactic disk. The relative shape of the thermal background was fixed to that fitted to the spectrum of an elliptical background region located outside the Vela SNR shell, centered on $(\alpha, \delta) = (08^{\rm h}44^{\rm m}30^{\rm s},-39^{\circ}26^{\prime})$, and with semi-axes of $2.6^{\circ}\times1.3^{\circ}$.    

\begin{figure*}[]
\centering
\includegraphics[width=18.0cm]{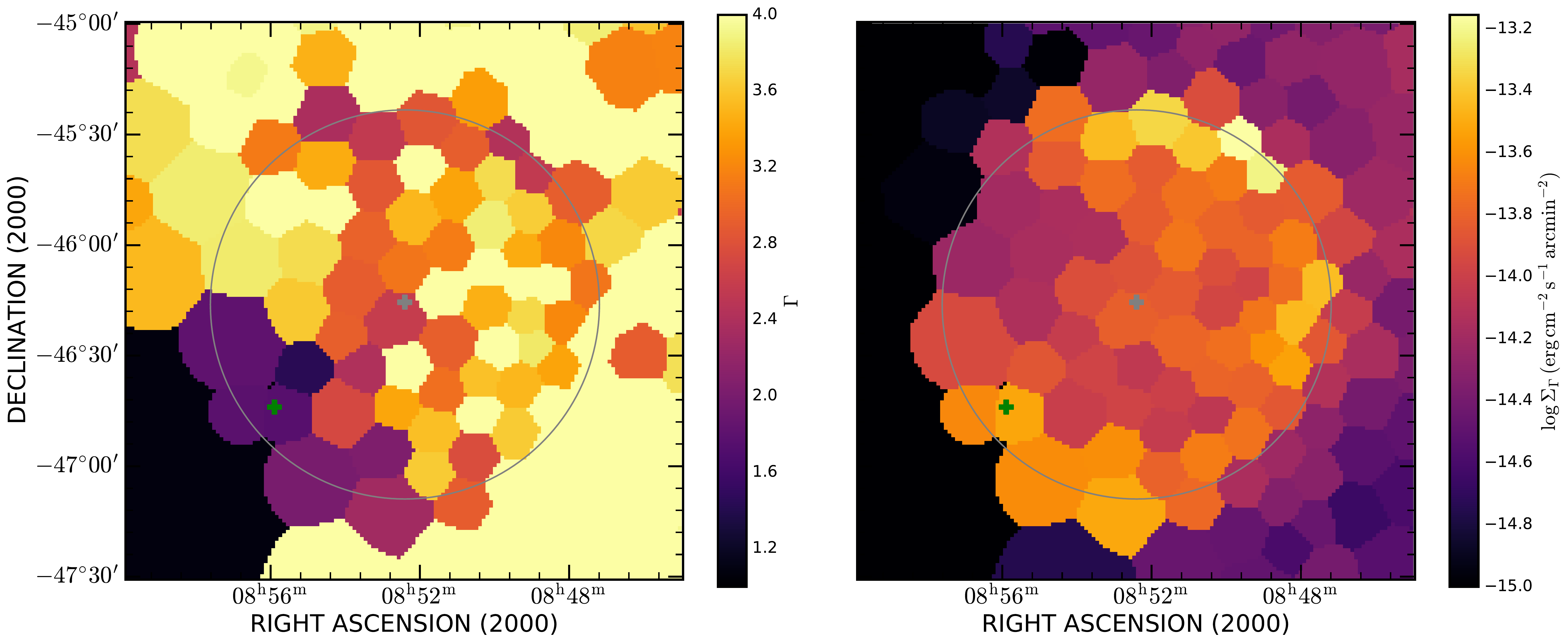} 
\includegraphics[width=18.0cm]{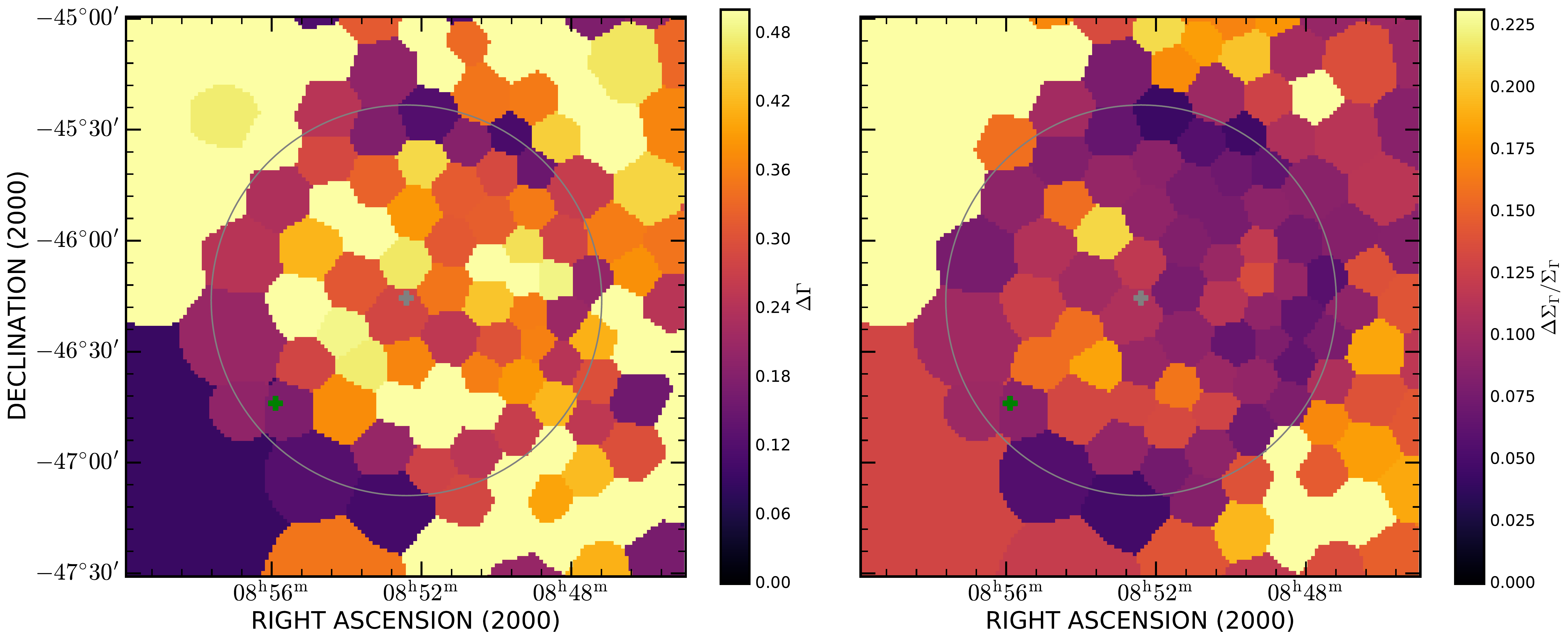} 
\caption{Properties of nonthermal emission across Vela Jr. 
The top panels illustrate the distribution of the photon index $\Gamma$ (left) and   the surface brightness $\Sigma_{\Gamma}$ (right), that is, the integrated flux per surface area of the nonthermal model components in spectral fits to Voronoi bins. The energy range for the computation of the integrated nonthermal flux is $1.0-5.0\,\mathrm{keV}$. 
The bottom   panels illustrate  the corresponding $1\sigma$ uncertainties of photon index (left) and surface brightness (right) measurements in each bin.
The gray circle gives the extent of Vela Jr, and is identical to the circle displayed in Figure~\ref{fig:geometric_center}. The position of the CCO and the energetic pulsar PSR J0855$-$4644, located in the southeast region of Vela Jr, are respectively indicated with a gray and green plus sign (+).} 
\label{VoronoiResults}
\end{figure*}

From the fit results, we created maps illustrating the spatial distribution of the best-fit physical parameters across Vela Jr, which are displayed in Figure~\ref{VoronoiResults}. 
%
%Figure 4
One can clearly see the signature of the Vela Jr shell in the map of the nonthermal brightness $\Sigma_{\Gamma}$ as it exceeds the local background outside its shell by around an order of magnitude. While the brightest emission is clearly detected along the northwestern, western, and southern rims, there is a significant spatially inhomogeneous excess of emission inside the shell. % when compared to the ``background'' level outside. %We interpret this as contribution from the Vela SNR (cf. also Figure~\ref{fig:VelaJr_sliced}).
We  note, however, that the map indicates a nonzero level of what appears to be nonthermal emission outside the shell of Vela Jr. We believe this is likely to be caused by the contribution of a hotter thermal emission component in the Vela SNR, and probably does not correspond to a truly nonthermal contribution. 
In fact, modeling the thermal background as a combination of two thermal components in equilibrium ({\tt vapec+vapec}) is a valid alternative approach \citep[see][]{LuAschenbach00}, which we tested. We recovered a very similar morphology for the nonthermal emission of Vela Jr, while the detected level of emission in the region outside was reduced. 
Importantly, even without considering alternate models, the measured level of nonthermal emission on the inside of the shell visible in Fig.~\ref{VoronoiResults} is significantly above the background level outside, especially in the western half. Thus, the presence of diffuse nonthermal emission on the inside of the shell of Vela Jr appears likely, whereas we do not claim to observe diffuse nonthermal emission from the Vela SNR. The properties of the suspected diffuse nonthermal component in Vela Jr are investigated in more detail in Sect.~\ref{sec:selected_regions}.       

The distribution of the photon index $\Gamma$ of the power law component appears to indicate comparatively hard nonthermal emission along the northwest and especially the southeast portions of the rim of Vela Jr. 
However, we believe that some skepticism is warranted regarding the physical origin of suspiciously small photon indices ($\Gamma \sim 1.4-1.8$) in the southeast. On one hand, visual inspection of selected spectral fits in this region shows that the detected nonthermal emission likely does have an astrophysical origin within the shell of Vela Jr, as its level is above that expected for nonthermal or instrumental backgrounds by a factor of a few. On the other hand, a possible issue with our approach is that we assume the absorption toward the entire Vela Jr remnant to be compatible with that of the CCO ($N_{\rm H} = 3.5\times10^{21}\,\mathrm{cm^{-2}}$). 
If the true absorption were much higher than that in the southeast, %on the level of $\sim 10^{22}\,\mathrm{cm^{-2}}$, 
the nonthermal component would experience significant downward curvature even above $1 \,\rm keV$, artificially reducing the inferred power law slope, in particular since the thermal component is relatively soft in this region. 
We tested this hypothesis by carrying out an array of simple XSPEC simulations using parameters for the thermal component typical for the southeast, but fixing the intrinsic power law photon index to $\Gamma = 2.5$ and varying the true absorption of the nonthermal component in the range $4-14\times10^{21}\,\mathrm{cm^{-2}}$. This showed that, when fitted with a fixed absorption as described above, the measured photon index reaches values of $1.8$ and $1.4$ at intrinsic column densities of $N_{\rm H} = 9\times10^{21}\,\mathrm{cm^{-2}}$ and $N_{\rm H} = 1.2\times10^{22}\,\mathrm{cm^{-2}}$, respectively. The physical origin of this suspected additional foreground absorption may be in the Vela molecular ridge (VMR), which spatially overlaps the southeastern part of the shell, if we assume that the shell of Vela Jr is located within or behind it, requiring a distance to Vela Jr $\gtrsim 700\,\mathrm{pc}$ \citep{1992A&A...265..577L}.  
%This may be the case in particular for the suspiciously hard photon indices $\Gamma \lesssim 2$ measured along the southeast rim, where the Vela molecular ridge (VMR) might contribute large foregroun d absorption, if the shell of Vela Jr is located within or behind it. 
For completeness, we note that the pulsar PSR J0855$-$4644 and its faint X-ray nebula overlap the southeastern rim as well \citep{2018MNRAS.477L..66M}. However, since the extent of this nebula is limited to less than an arcminute in the \textit{Chandra} data \citep{2017A&A...597A..75M}, and no extended emission from it is visible in the eRASS:4 intensity image (see Fig.~\ref{fig:VelaJr_sliced}), it seems unlikely to be 
responsible for the hard emission detected in the southeast rim \citep[see][for a discussion of   the pulsar and of the association with the VMR]{2013A&A...551A...7A}. 
% Furthermore, a contribution of hard photons from PSR J0855$-$4644 and its nebula \citep{2018MNRAS.477L..66M} may be relevant here too  \citep[see][for a discussion of both the pulsar and the association with the VMR]{2013A&A...551A...7A}. 

\subsubsection{Spectral modeling of selected regions} \label{sec:selected_regions}
\begin{figure}
    \centering
    \includegraphics[scale=0.35]{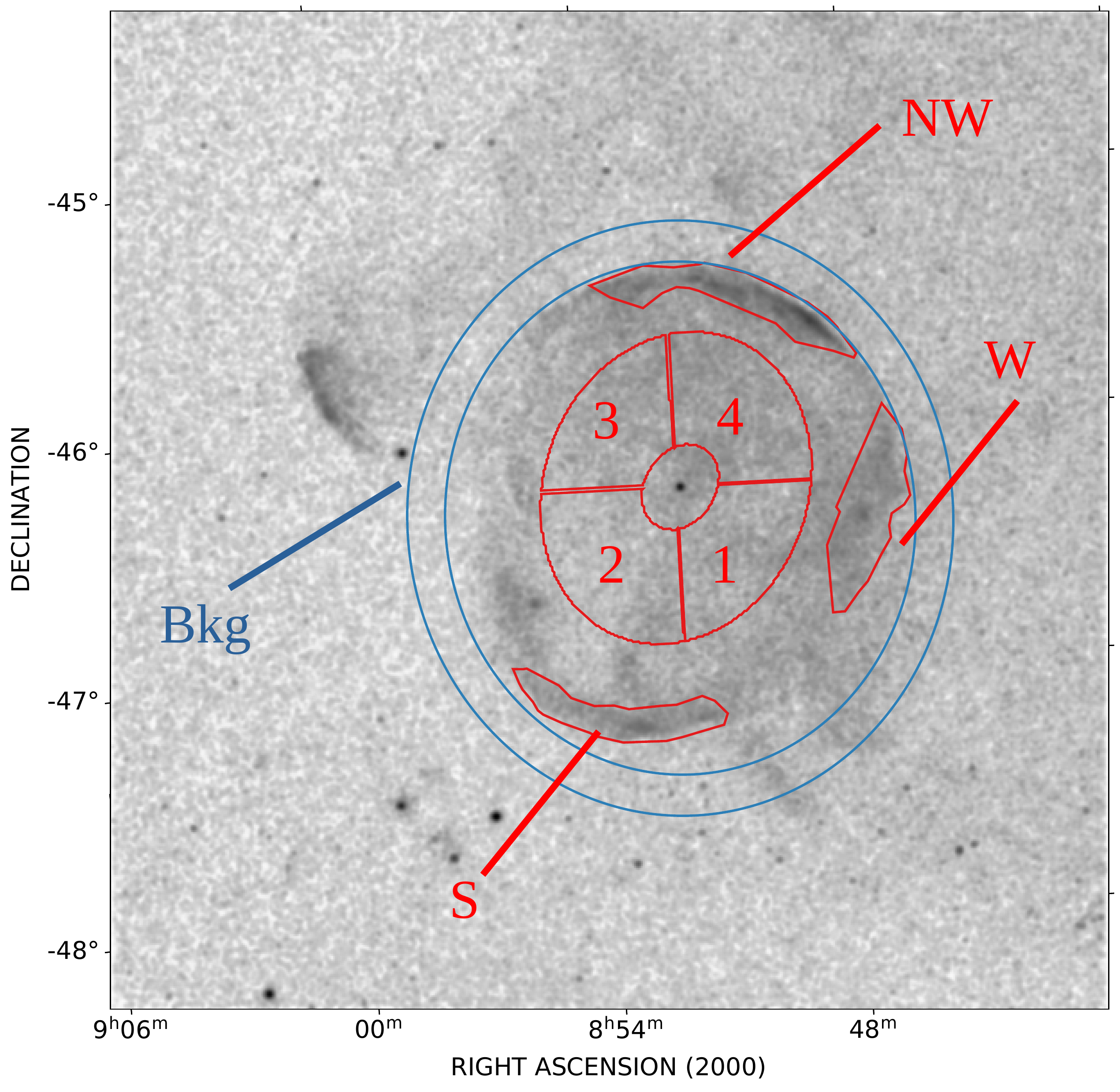}
    \caption{Vela Jr and the surrounding region in the $1-8$ keV band, taken from stacked eRASS:4 data. Source regions selected for spectral analysis are highlighted in red, whereas the region  used to model the background spectrum is indicated in blue.
    \label{fig:VelaJr_regions}}
\end{figure}

Due to the limited photon statistics  in the eROSITA survey data of Vela Jr, we divided 
the entire remnant into seven distinct regions, which allowed us to obtain spectral 
parameters with higher accuracy than is possible in the Voronoi binning approach. 
As indicated in Figure \ref{fig:VelaJr_regions}, the defined areas include three boundary 
shock regions and four sectors located well inside the remnant. 
%
%Figure 5
From Figures \ref{fig:VelaJr_RGB}  and \ref{fig:VelaJr_sliced} it is clear how the contribution of the Vela SNR is dominant at energies below $\sim 1$ keV and how Vela Jr contributes to the emission above this energy. In a first attempt, we therefore tried to restrict our spectral 
analysis to the $1-5$ keV energy band to exclude most of the contribution from the Vela SNR. However, after a detailed analysis, we concluded that this interval did not allow us to 
fully constrain all the physical spectral components. Finally, we used the full $0.3-10$ keV
bandwidth for the spectral analysis and fitted the contribution from the Vela SNR first 
in order to get a background model. Building up on that, we fitted the source spectra for the distinct regions of Vela Jr simultaneously with the background model. It is worth  noting that the background was taken from an elliptical ring surrounding Vela Jr so that the background
model obtained this way may be only an approximation of the real Vela SNR contribution for
regions within Vela Jr. Details of the background modeling and spectral fitting are described in  Appendix \ref{app:eRASS}. 

Starting from these, we employed an approach based on the Markov chain Monte Carlo (MCMC) technique to have a better understanding of the correlation between the various spectral parameters of the tested models. This delivers a more robust error estimation \citep[see, e.g.,][respectively for a review of the method and an applied case to SNR]{vanDyk2001,Borkowski2010}. We ran the Python library \texttt{emcee}-based code \citep{2013PASP..125..306F}   with 40000 steps to make sure the chains could converge. We then manually inspected the trace plot of the chains, selecting only the last 2000 steps. This choice ensured that most of them were converged. Even though MCMC-based approaches are not designed to give a quantitative estimate of the goodness of the fit (unlike the $\chi^{2}$ estimator), we present in Table \ref{tab:BF_statistic} the values of the median Cash statistic and degree of freedom values drawn from the converged MCMC chains. In this sense, we employ the ratio CSTAT/dof and also the reduced $\chi^{2}/dof=\chi^{2}_{dof}$. 

% Table 2
\begin{table*}[]
\caption{Best-fit parameters for the model VPSHOCK. In the upper part are reported the parameters of the source, while in the lower part   those from the background.\label{tab:BF_statistic}}
\begin{tabular}{ccccccccc}
\hline
& & & VPSHOCK &&& & &\\[1ex]%
\hline
&NW&S&W&sector1&sector2&sector3&sector4&\\[1ex]%
\hline%
Source model& & & &&& & &\\[1ex]%
\hline%
factor&$0.0447_{-0.0017}^{+0.0016}$&$0.0393_{-0.0016}^{+0.0023}$&$0.0488_{-0.0016}^{+0.0018}$&$0.0760_{-0.0018}^{+0.0022}$&$0.078_{-0.002}^{+0.002}$&$0.080_{-0.002}^{+0.003}$&$0.085_{-0.002}^{+0.002}$\\[1ex]%
N$\_$H (10$^{22}$ cm$^{-2}$)&$0.47_{-0.10}^{+0.13}$&$0.44_{-0.08}^{+0.11}$&$0.004_{-0.002}^{+0.009}$&$0.2_{-0.2}^{+0.6}$&$0.004_{-0.002}^{+0.009}$&$0.004_{-0.002}^{+0.011}$&$1.1_{-0.3}^{+0.3}$\\[1ex]%
kT (keV)&$1.40_{-0.15}^{+0.22}$&$1.9_{-0.3}^{+0.4}$&$1.9_{-0.2}^{+0.2}$&$0.83_{-0.07}^{+0.06}$&$9.1_{-1.0}^{+0.7}$&$0.30_{-0.03}^{+0.02}$&$1.0_{-0.3}^{+0.3}$\\[1ex]%
C/C$_{\odot}$&$0.02_{-0.02}^{+0.22}$&$0.1_{-0.0}^{+0.6}$&$0_{-0}^{+4}$&$0.1_{-0.1}^{+2.0}$&$0.007_{-0.005}^{+0.032}$&$1.0_{-0.5}^{+1.1}$&$0_{-0}^{+4}$\\[1ex]%
N/N$_{\odot}$&$0.01_{-0.01}^{+0.17}$&$0.01_{-0.01}^{+0.07}$&$0.1_{-0.1}^{+1.5}$&$0.1_{-0.1}^{+1.6}$&$0.008_{-0.006}^{+0.031}$&$0.2_{-0.2}^{+0.4}$&$0.2_{-0.1}^{+2.9}$\\[1ex]%
O/O$_{\odot}$&$0.02_{-0.02}^{+0.09}$&$0.03_{-0.03}^{+0.09}$&$0.01_{-0.01}^{+0.06}$&$3_{-2}^{+4}$&$0.45_{-0.03}^{+0.03}$&$1.3_{-0.3}^{+1.3}$&$3_{-2}^{+3}$\\[1ex]%
Ne/Ne$_{\odot}$&$0.27_{-0.12}^{+0.16}$&$0.10_{-0.07}^{+0.15}$&$0.1_{-0.1}^{+0.3}$&$0.1_{-0.1}^{+0.6}$&$0.88_{-0.06}^{+0.06}$&$2.1_{-0.5}^{+1.6}$&$1.2_{-0.6}^{+0.7}$\\[1ex]%
Mg/Mg$_{\odot}$&$0.17_{-0.16}^{+0.17}$&$0.03_{-0.03}^{+0.15}$&$0.01_{-0.01}^{+0.07}$&$0.5_{-0.3}^{+0.6}$&$0.53_{-0.06}^{+0.08}$&$1.9_{-0.5}^{+1.8}$&$0.31_{-0.21}^{+0.19}$\\[1ex]%
Si/Si$_{\odot}$&$0.4_{-0.4}^{+0.4}$&$0.04_{-0.04}^{+0.24}$&$0.05_{-0.04}^{+0.15}$&$0.6_{-0.3}^{+0.4}$&$0.25_{-0.15}^{+0.15}$&$0.3_{-0.3}^{+1.1}$&$0.02_{-0.02}^{+0.08}$\\[1ex]%
Fe/Fe$_{\odot}$&$0.01_{-0.01}^{+0.14}$&$0.02_{-0.02}^{+0.10}$&$0.007_{-0.005}^{+0.024}$&$0.3_{-0.2}^{+0.2}$&$0.008_{-0.006}^{+0.031}$&$0.37_{-0.09}^{+0.18}$&$0.01_{-0.01}^{+0.13}$\\[1ex]%
Tau$_u$ (10$^{10}$ cm$^{-3}$ s)&$0.10_{-0.05}^{+0.05}$&$0.09_{-0.06}^{+0.19}$&$400_{-200}^{+400}$&$200_{-200}^{+400}$&$0.87_{-0.08}^{+0.10}$&$130_{-50}^{+140}$&$0_{-0}^{+400}$\\[1ex]%
Normalization&$0.66_{-0.10}^{+0.13}$&$0.31_{-0.04}^{+0.07}$&$0.119_{-0.008}^{+0.008}$&$0.052_{-0.014}^{+0.025}$&$0.043_{-0.003}^{+0.003}$&$0.16_{-0.10}^{+0.04}$&$0.06_{-0.02}^{+0.05}$\\[1ex]%
\hline%
Background model& & & &&& & &\\[1ex]%
\hline%
N$\_$H (10$^{22}$ cm$^{-2}$)&$0.008_{-0.005}^{+0.006}$&$0.004_{-0.003}^{+0.010}$&$0.009_{-0.002}^{+0.003}$&$0.014_{-0.007}^{+0.007}$&$0.003_{-0.002}^{+0.003}$&$0.003_{-0.001}^{+0.004}$&$0.021_{-0.009}^{+0.003}$\\[1ex]%
kT (keV)&$0.29_{-0.03}^{+0.03}$&$0.24_{-0.02}^{+0.02}$&$0.310_{-0.009}^{+0.008}$&$0.27_{-0.02}^{+0.03}$&$0.327_{-0.015}^{+0.013}$&$2.1_{-1.3}^{+1.0}$&$0.206_{-0.005}^{+0.018}$\\[1ex]%
Tau$_u$ (10$^{10}$ cm$^{-3}$ s) &$17_{-4}^{+9}$&$35_{-10}^{+22}$&$19_{-2}^{+3}$&$23_{-7}^{+10}$&$20_{-3}^{+4}$&$0.3_{-0.1}^{+2.4}$&$90_{-30}^{+20}$\\[1ex]%
Normalization&$0.30_{-0.05}^{+0.06}$&$0.37_{-0.04}^{+0.11}$&$0.301_{-0.018}^{+0.021}$&$0.35_{-0.06}^{+0.08}$&$0.268_{-0.017}^{+0.021}$&$0.05_{-0.01}^{+0.05}$&$0.67_{-0.13}^{+0.06}$\\[1ex]%
\hline
CSTAT/dof & 1.10 &1.20 & 1.19&1.13&1.25&1.29&1.29\\[1ex]%
\end{tabular}
\end{table*}

The first model tested for each of the seven regions was VPSHOCK, which represents a single temperature plane-parallel shocked plasma with non-equilibrium collisional ionized material \citep{2001ApJ...548..820B}, allowing us to fit the single abundance parameters. We left free to vary C, N, O, Ne, Mg, Si, and Fe.  We also tested the power law model in order to compare a featureless model spectrum against a thermal one. Both models included Galactic absorption using the TBabs model.

Starting from the NW rim region, we fitted a dominating nonthermal model to the featureless spectrum above 1 keV. Comparing the results of the different models, the reduced statistic values of VPSHOCK and power law are identical ($\chi^{2}/$dof=1.1, see Table \ref{tab:BF_statistic}) . To assess which was the most reliable model, we compared the corner plots given by the two models, concluding a power law model gives more robust constrained parameters. Our preferred conclusion is that the spectrum is featureless and a power law is the best-fitting model for it. 

\begin{figure*}
     \centering    \includegraphics[width=6cm]{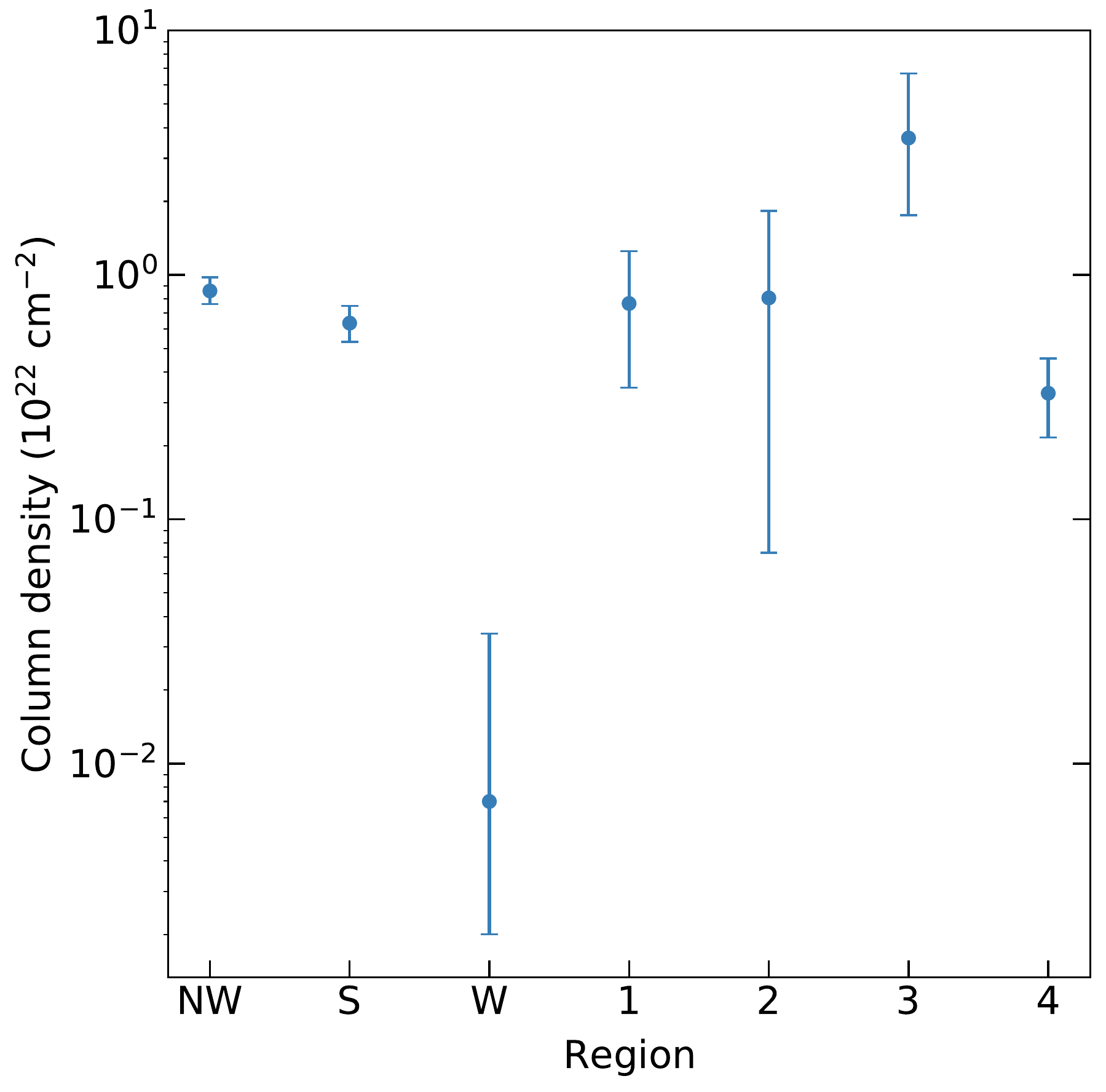}
     \includegraphics[width=5.6cm]{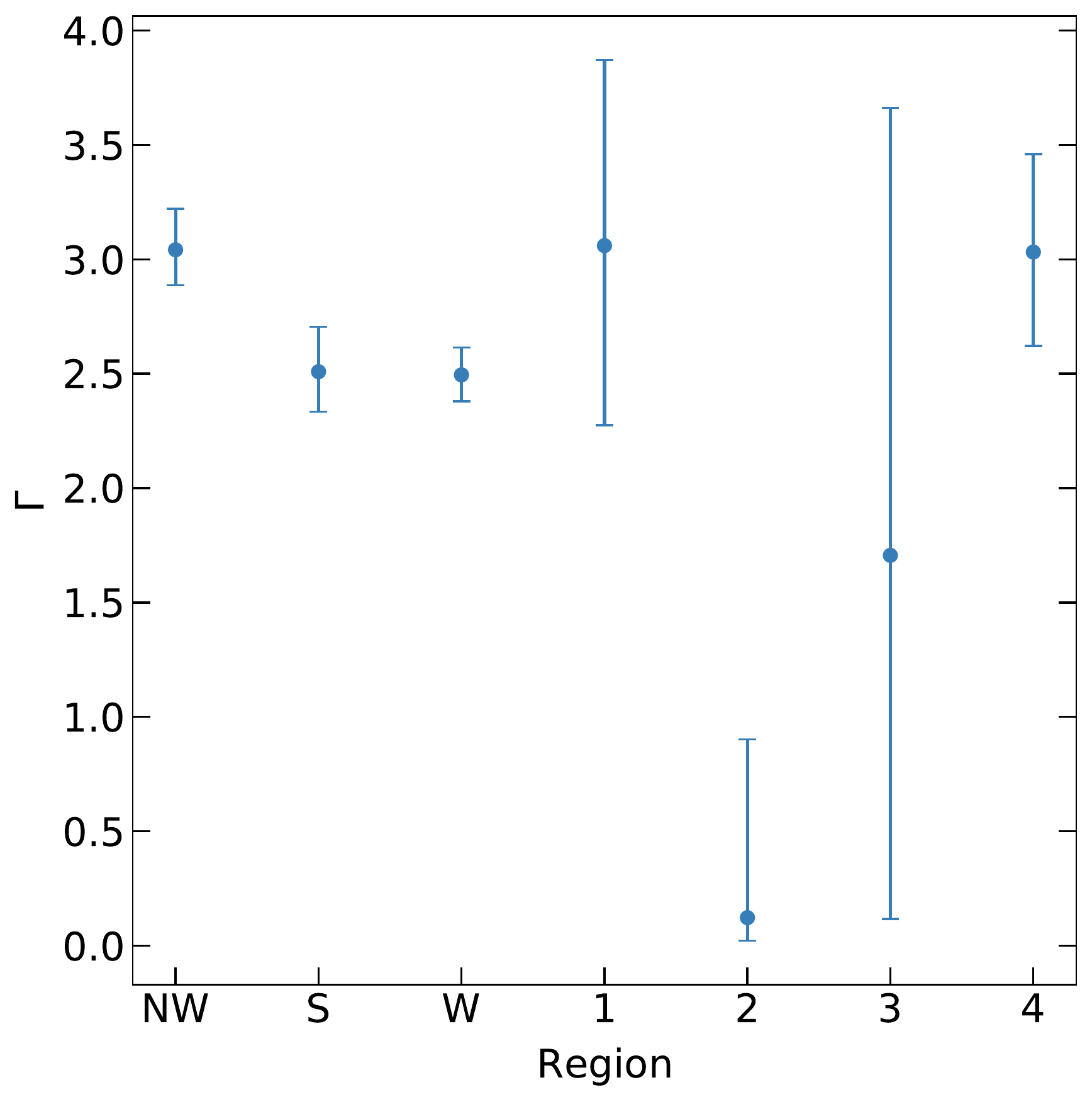}
     \includegraphics[width=6cm]{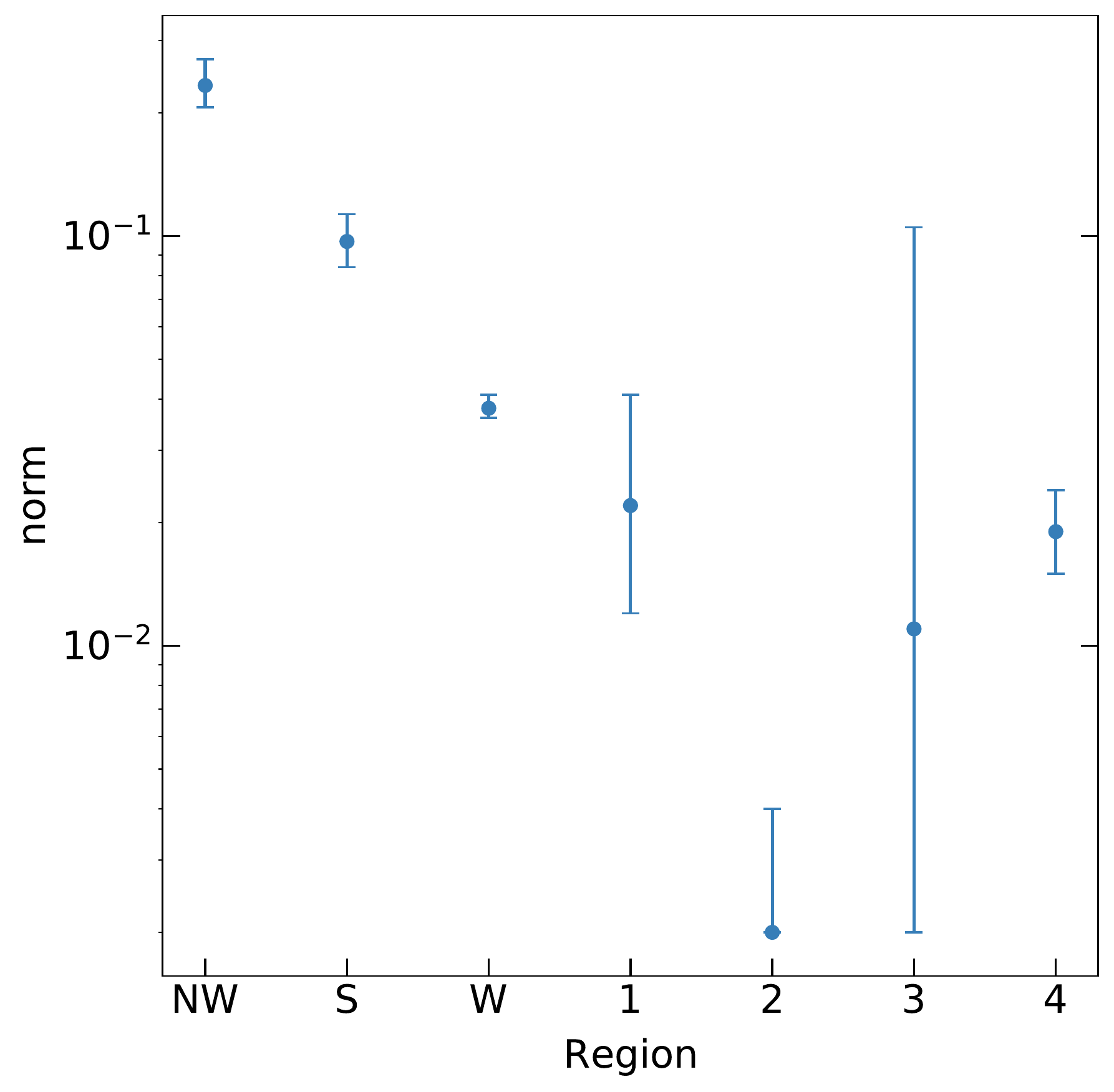}
    \caption{Distribution of spectral parameters for the power law model when fitted to the spectra of the selected regions defined in Figure \ref{fig:VelaJr_regions}. The photon index $\Gamma$ plot clearly shows how sectors 1 and 4 have  well constrained values, similarly to the NW, W, and S regions, while sectors 2 and 3 do not.
    \label{fig:eRASS1234_powerlaw_parameters}}
\end{figure*}

\begin{figure*}
     \centering    \includegraphics[width=6cm]{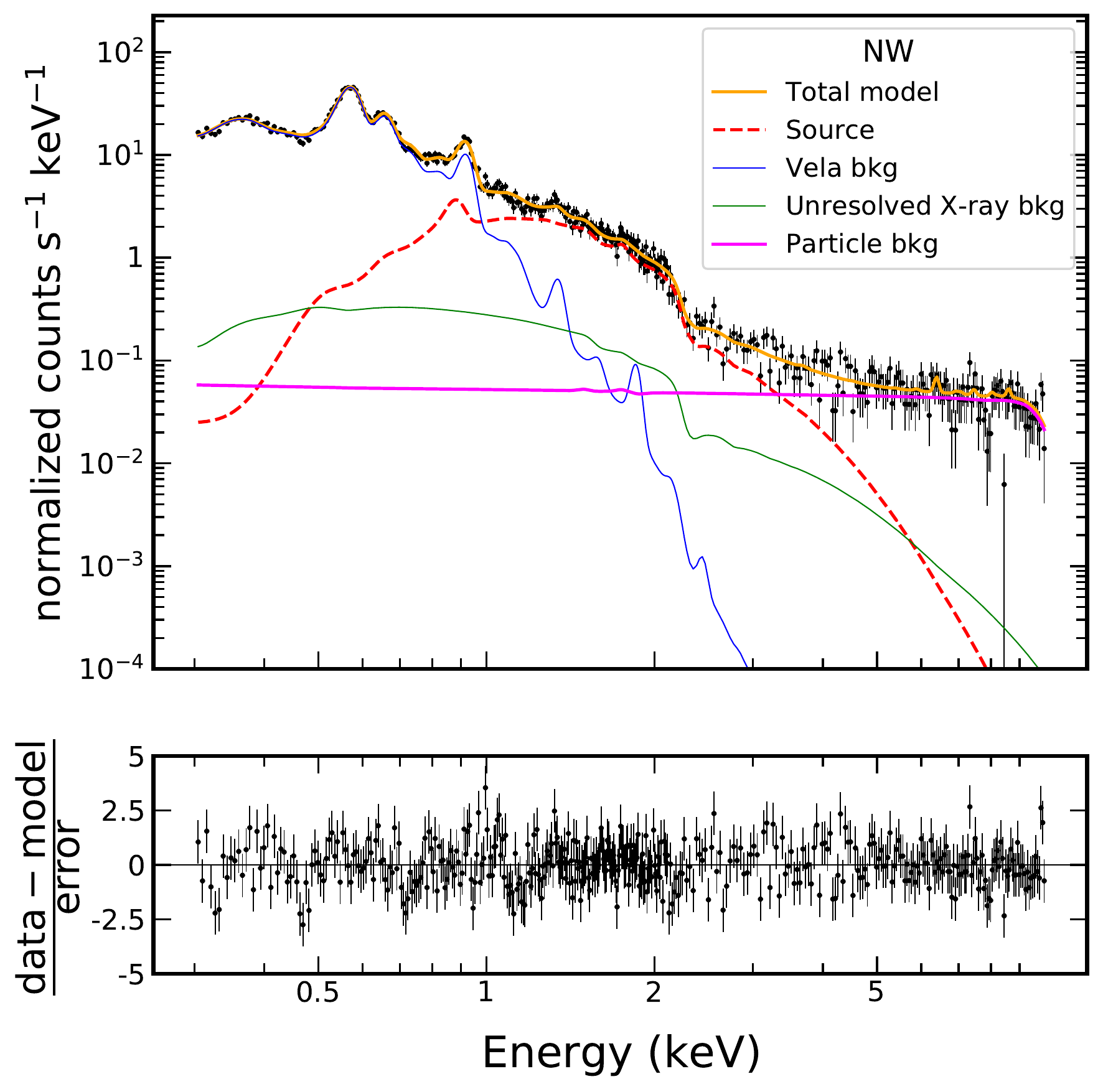}
     \includegraphics[width=6cm]{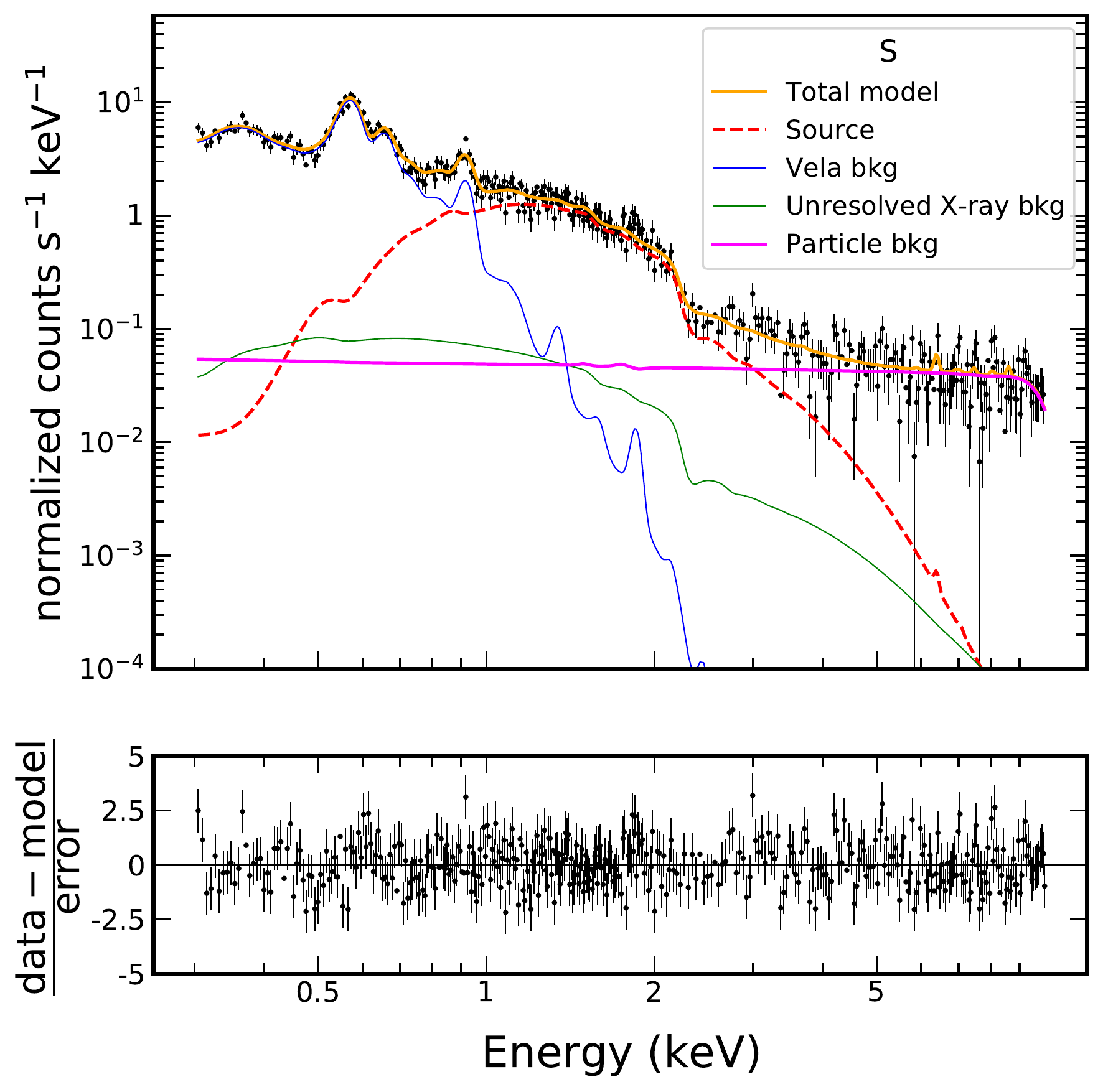}
     \includegraphics[width=6cm]{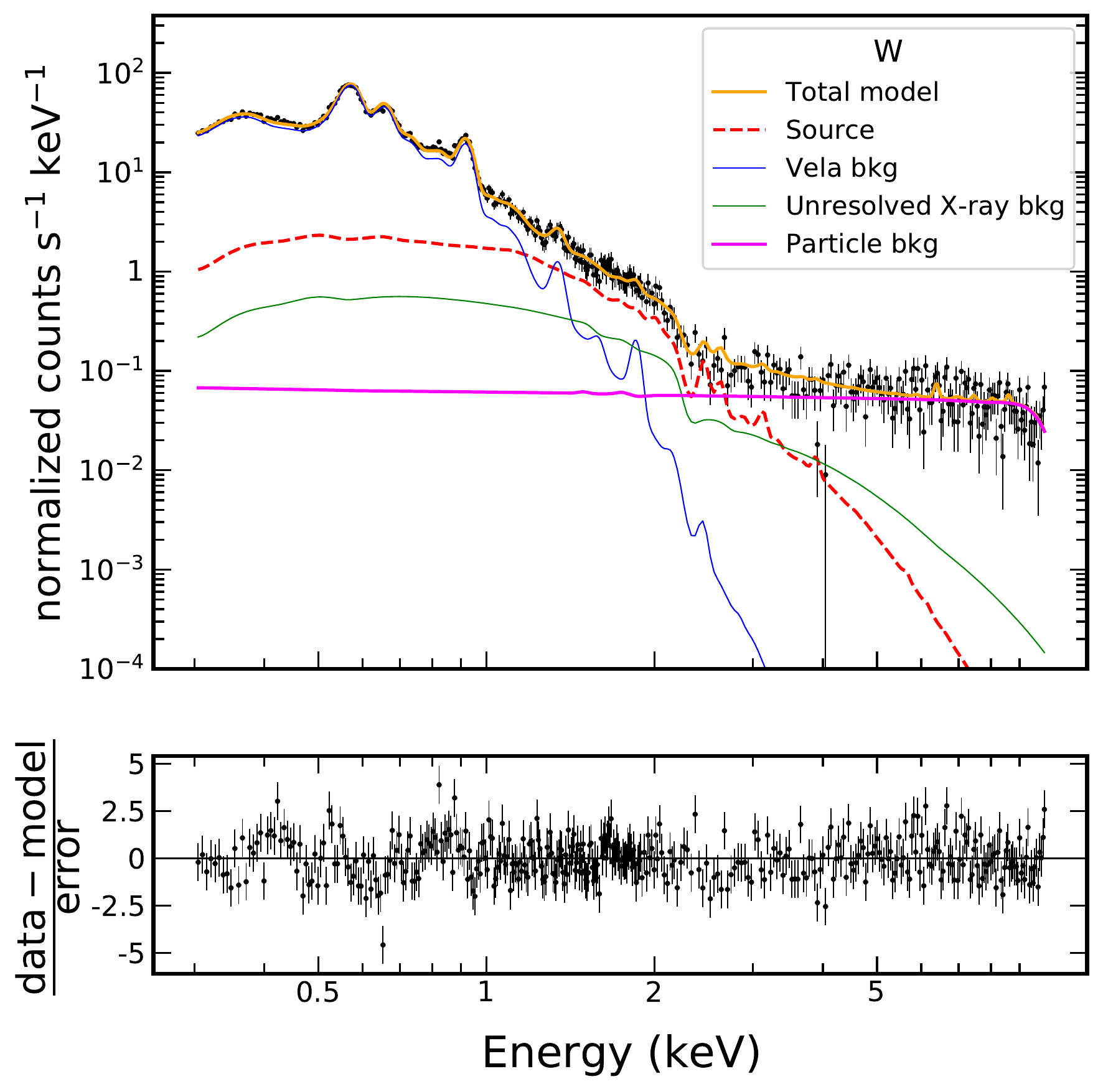}
     \includegraphics[width=6cm]{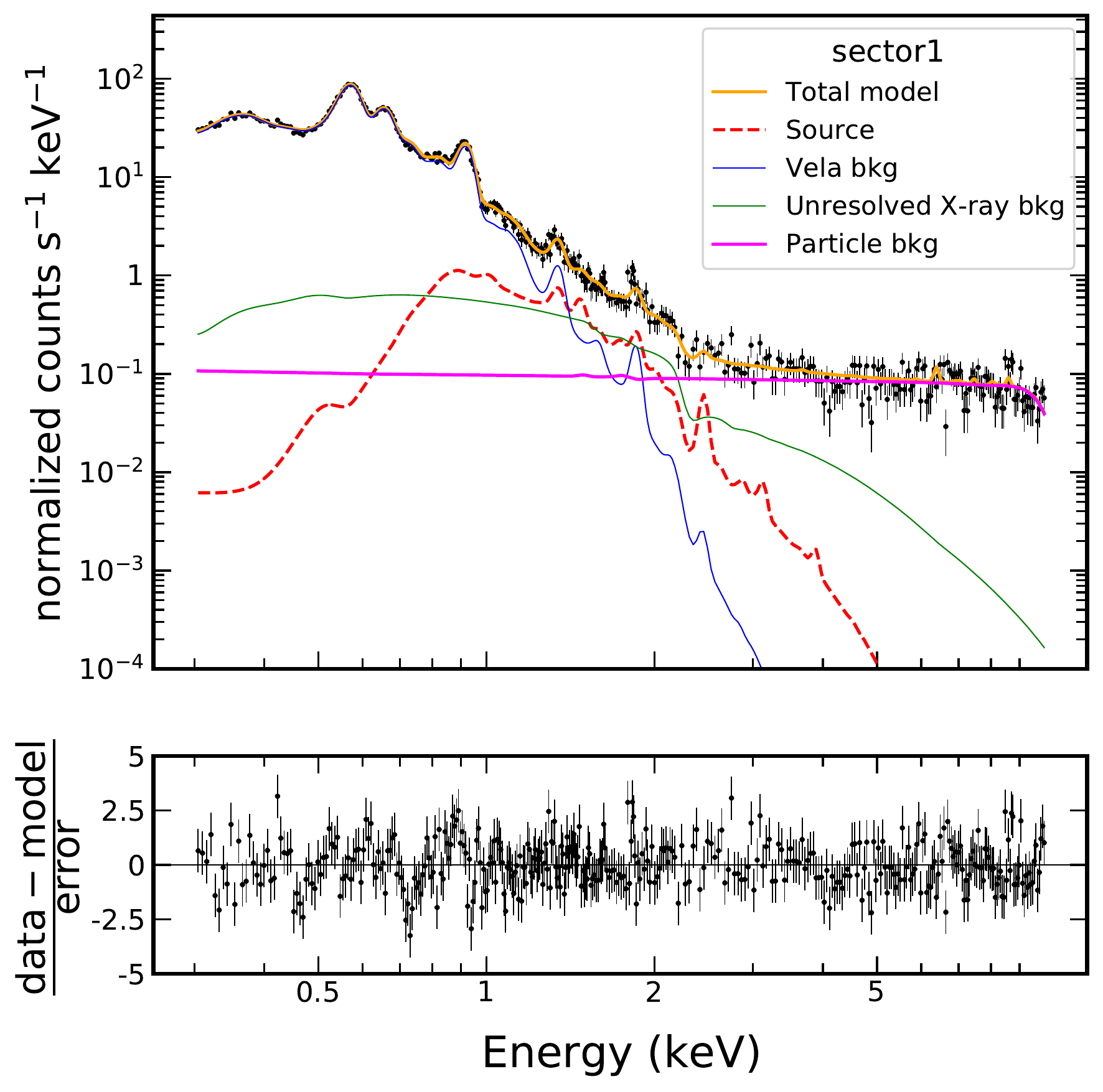}
     \includegraphics[width=6cm]{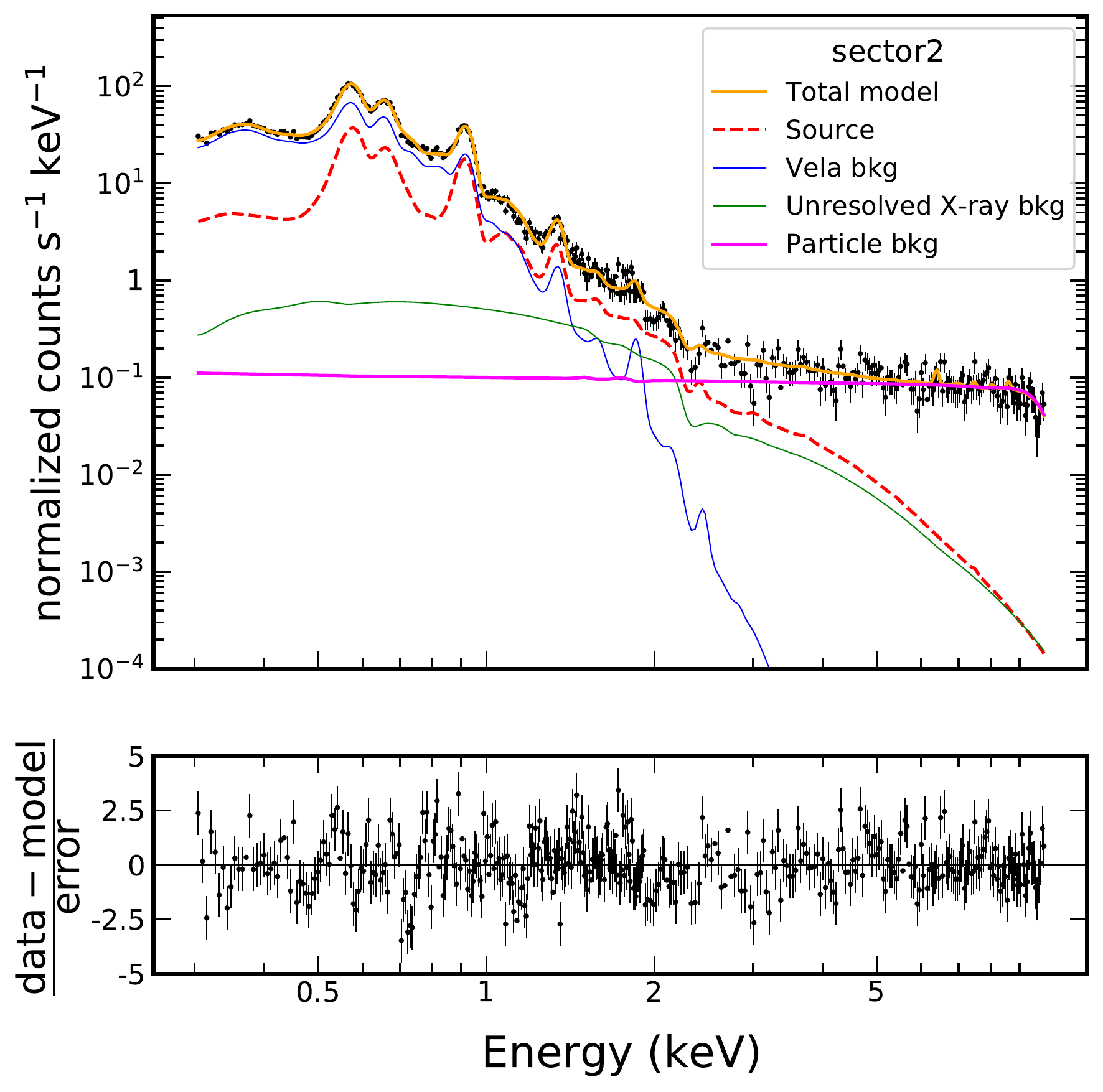}
     \includegraphics[width=6cm]{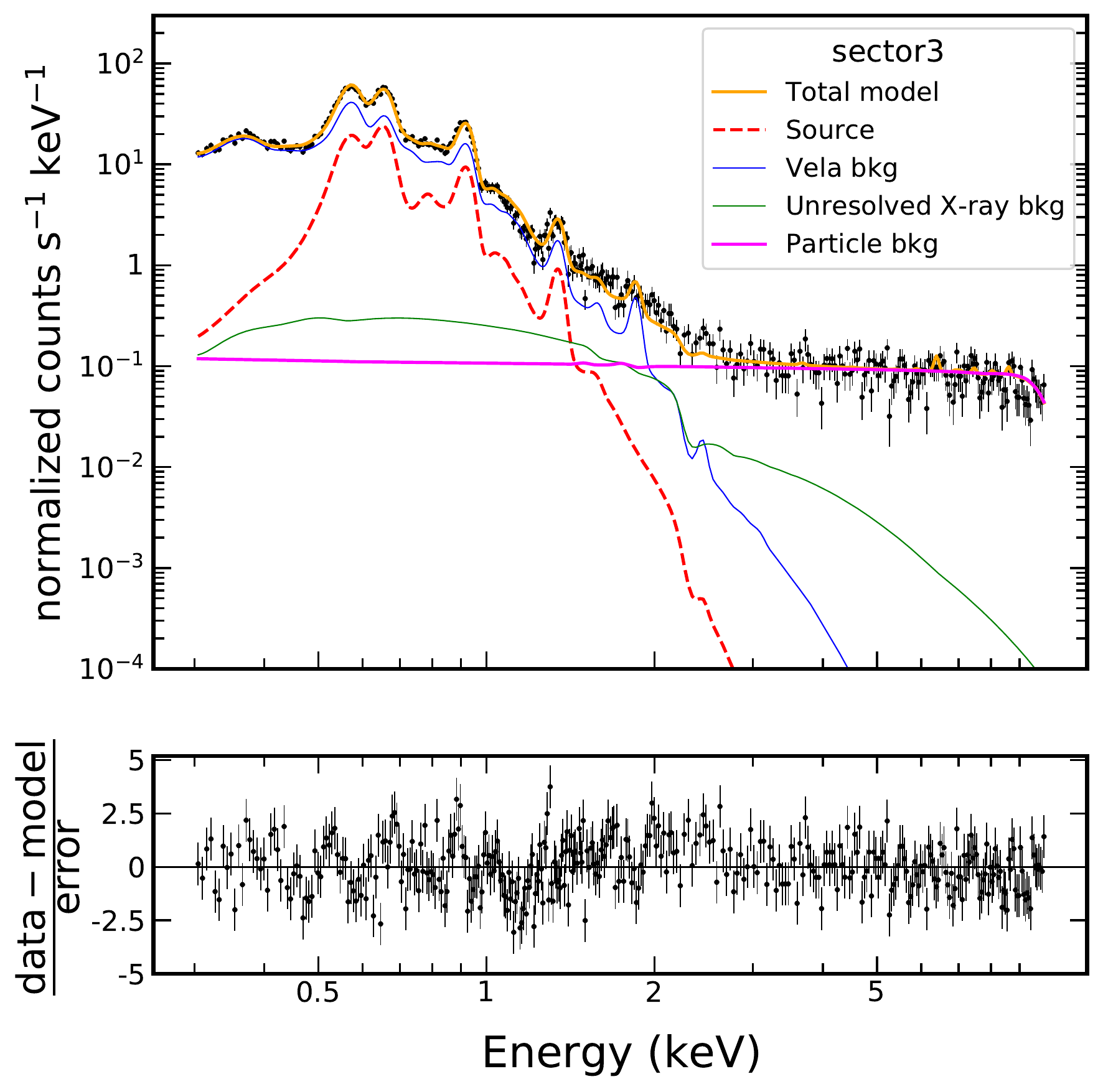}
     \includegraphics[width=6cm]{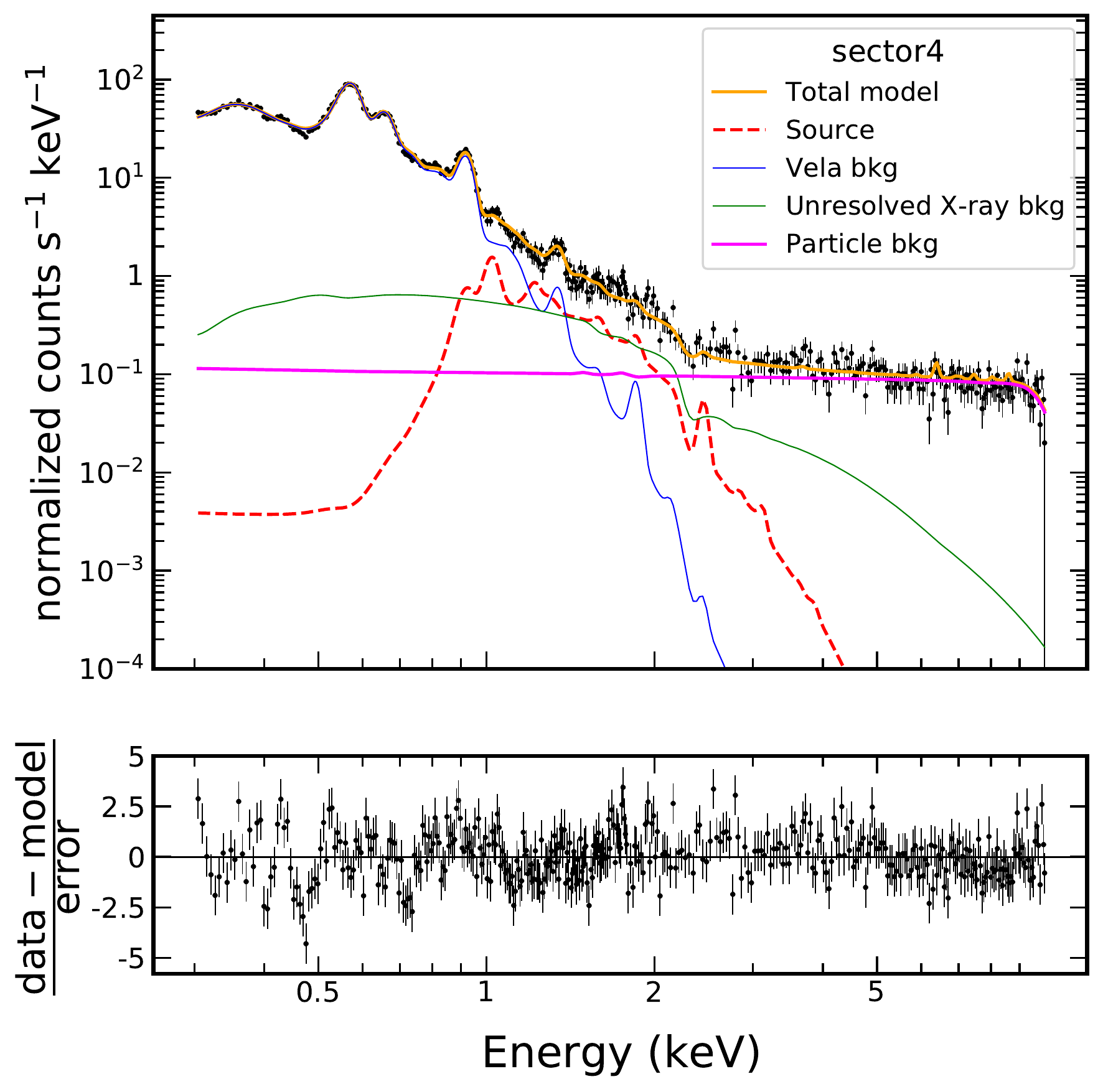}
    \caption{eRASS:4 spectra (from top left to bottom: NW, S, W, sector 1, sector 2, sector
3, and sector 4)  fitted with non-equilibrium parallel shock thermal model (VPSHOCK). Red  indicates the source model, green and blue represent the components of the X-ray background model, and  magenta indicates the most significant component of the particle background. The spectra were rebinned for graphical purposes.
    \label{fig:eRASS1234_vpshock_whole}}
\end{figure*}

Regarding the west region (labeled  ``W'' in Figure \ref{fig:VelaJr_regions}), it seems to be dominated by nonthermal emission as well, showing no strong lines above 1 keV. However, comparing it to the NW rim region, the spectrum of the W rim appears flatter. Using a power law model, we observe $\Gamma =3.05 \pm 0.17$ at the NW rim against $\Gamma =2.5 \pm 0.12$ of the W rim obtained with the same model (Figure \ref{fig:eRASS1234_powerlaw_parameters}). We further observed that the W rim has a lower value for the column density than those of other regions, while the normalization and photon indices are more similar to each other. Another featureless spectrum is obtained from the S rim. The best-fitting model is clearly a power law for this region (cf. Figure \ref{fig:eRASS1234_vpshock_whole}). The photon index and column density values for that region are indeed similar to those of the NW rim. 

%Figure 6

The four inner regions of the remnant are all characterized by the presence of several unresolved lines above 1 keV, while the outer regions, for example  NW, W, and S, show featureless spectra at this energy range. Sectors 2 and 3 are basically dominated by background emission (which is emission from the Vela SNR). The VPSHOCK and power law model fits for sectors 1 and 4 have a comparable goodness (see Table \ref{tab:BF_statistic}),
whereas for sectors 2 and 3 the spectral fits are definitely worse when a power law model is fitted.

Regarding sectors 2 and 3, we tested a two-component model made by a VPSHOCK plus a power law component (i.e., a thermal plus a non-thermal component). In both regions we obtain good fitting results; the extra thermal component consistently improves the fit, compared to the single power law  model (see Table \ref{tab:BF_statistic}). If in sector 2 the power law photon index is set to a value close to those of the other regions ($\Gamma \sim 2.5$) then the thermal component is largely unconstrained. Conversely, the thermal component is stronger in sector 3 and the power law is not well constrained. However, a combination of the two is always needed to improve the fit statistic. We note again how these spectra suffer from degeneracy of several factors, mainly the superposition of the Vela SNR and Vela Jr along the line of sight. In conclusion, the spectrum of sector 2 is probably featureless, as are those of sector 1 and 4, while in sector 3 an extra thermal component (probably due to  the incomplete background modeling of the Vela SNR contribution) is needed to have a good fit. In Figure \ref{fig:eRASS1234_powerlaw_parameters}, it is shown how the boundary regions selected in NW, W, and S have strong nonthermal spectra. Therefore we conclude that the spectrum of Vela Jr is uniformly nonthermal across the whole remnant. 
\subsection{eROSITA CalPV and \textit{XMM-Newton} data of the NW rim \label{sec:NW_only}}

\begin{figure}[h]
    \centering
    \includegraphics[scale=0.323]{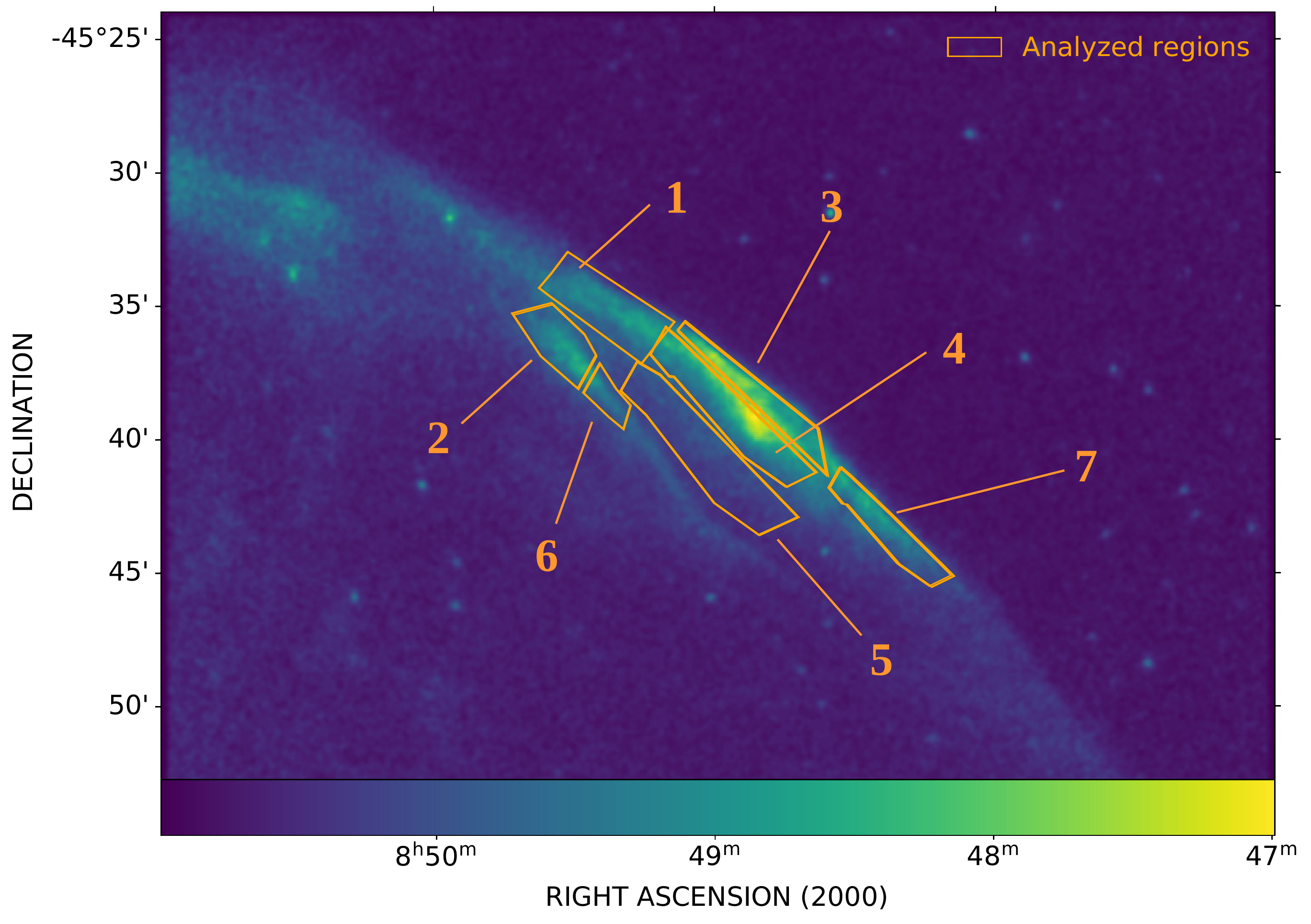}
    \caption{Exposure- and vignetting-corrected eROSITA image of the 
    NW rim of Vela Jr, restricted to the $1.1-8.0$ keV energy band. The 
    labeled subregions indicate the sectors whose spectra were 
    extracted. The image pixel size is 4" and a Gaussian smoothing 
    with $\sigma=1.3$ pixels has been applied. The color scale ranges 
    from zero (dark blue) to a maximum of 0.0011 counts/s (yellow). \label{fig:eROSITA_CalPV_image_b}}
\end{figure}

In addition to the survey observations described in the previous section, eROSITA observed  the NW rim of Vela Jr during its CalPV phase in a 60 ks deep pointed
observation.  \textit{XMM-Newton} has also observed the NW rim of Vela Jr on a regular basis
for the purpose of instrument calibration. In total, we found six data sets in the
public archive (see Table \ref{tab:observations}) taken between 2001 and 2021, which
sum up to a total exposure  of  more than 410 ks. We used this eROSITA
CalPV and \textit{XMM-Newton} archival data to search for temporal spectral changes in the
emission properties of the NW rim and to further constrain the results
obtained from eROSITA's survey data. The much higher photon statistics of this
data set allowed us to apply a spectral analysis for seven distinct regions of
the NW rim (see Figure \ref{fig:eROSITA_CalPV_image_b}). These regions were
tailored according to \cite{2010ApJ...721.1492P} in order to make the results
comparable with the one obtained from a \textit{Chandra} observation of that spot. The
background spectra for XMM and the CalPV data were extracted from an area north 
of the NW rim, located at RA=08h48m42s, DEC=$- 45^{\circ}$ $30$' $18$" and covering an elliptical
area of approximately $33 \times  11$ arcmin. 

%Figure 7

 The \textit{XMM-Newton} observations were taken in calibration mode, except the one from
 2001. This mode can create two disturbing effects in the data. The first is an
 unnatural rise in the continuum spectrum due to Bremsstrahlung produced by the 
 production of Auger electrons from the onboard calibration source. Both the source and
 background spectra are affected by that, but the effect is not homogeneous over the
 detector. The second effect is an uneven illumination of the PN and MOS1/2 detectors, making the
 background even more varying. After a careful background modeling (see Appendix
 \ref{app:XMM}), we did not observe any significant temporal variation in the spectra
 extracted from each of the six archival \textit{XMM-Newton} data sets taken between 2001 
 and 2021. This allowed us for each of the seven regions to fit the spectra from all
 \textit{XMM-Newton} observations simultaneously. The energy band was set to $0.3-10$ keV. From 
 this analysis, we obtained the best-fit value for the photon index and the column density
 of each of the seven regions (Table \ref{tab:NW_only_XMM}), averaged over six different observations. We then applied the MCMC approach as in Section \ref{sec:selected_regions}.
 For the \textit{XMM-Newton} data set we used 5000 steps because of the  very high photon 
 statistics available, whereas for the eROSITA CalPV data we had to apply 10000 steps 
 because of the  lower photon statistics.

 Based on the results obtained in the previous sections, we tested an absorbed power law model first. In order to correctly model the absorption, we employed the Tuebingen-Boulder model (TBABS in XSPEC; \citealt{2000ApJ...542..914W}). The fit gave well-constrained parameters in all regions. Based on the \textit{XMM-Newton} data, the fitted column density is found to be in the range $\simeq 2-8 \cdot 10^{21}$ cm$^{-2}$ (Table \ref{tab:NW_only_XMM}). With an uncertainty of typically $0.1 \cdot 10^{21}$ cm$^{-2}$ there is just a small variation in the column absorption among the seven defined regions of the NW rim. The fitted photon indices show a similar soft variation and are fitted in the range $2.6 - 3.1$. We ran the same fit on the CalPV data and found good agreement for all regions, except region 5. Comparing Table \ref{tab:NW_only_XMM} and Table \ref{tab:NW_only_CalPV} it seems CalPV derived parameters are systematically higher by approximately 20\% for regions 1, 2, and 3 compared to those derived with \textit{XMM-Newton}. These differences are easily explained with the early calibration status of eROSITA.

 After testing a power law, we tried a more complex model in order to characterize the underlying electron population. The SRCUT model \citep[see][for a comprehensive discussion]{1996ApJ...459L..13R,1998ApJ...493..375R}, which is implemented in XSPEC \citep{1996ASPC..101...17A}, convolves the synchrotron volume emissivity of a single electron with an exponentially cut off power law electron distribution. In other words, this model sets an upper limit to the cutoff energy of the electrons, without further physical assumptions \citep[for a detailed description of this model, see also][]{1999ApJ...525..368R}. The SRCUT model takes as parameter the radio spectral index $\alpha$ ($f(\nu)\propto \nu^{-\alpha}$, where $f$ is the radio flux) and the normalized radio flux at 1 Jy for the normalization. Therefore, in order to exclude a degeneracy between the parameters, one has to rely on radio measurements (either flux or radio spectral index) to constrain the break frequency. 
 We thus fixed the radio spectral index to -0.4 (see, e.g.,  the discussion in \citealt{2018ApJ...866...76M}) and left the normalization free to vary. As with the power law model, for each region we loaded all the
 spectra from the different observations and fitted them together. The background was handled
 as described in Appendix \ref{app:XMM}. In all regions we find a cutoff frequency consistent
 with 10$^{16}$ Hz. Following \cite{2010ApJ...721.1492P}, we can convert that frequency
 (parameter ``break'' in SRCUT) to the actual cutoff energy of the electrons according 
 to
\begin{equation}
\nu_{\text{cutoff}} \approx 1.6 \times 10^{16} \left(\dfrac{\text{B}}{\text{10 }\mu \text{G}}\right)\left(\dfrac{\text{E}_{\text{cutoff}}}{\text{10 TeV}}\right)^{2}\text{ Hz},   \label{eq:E_cutoff}
\end{equation}
where $\text{B}$ is the magnetic field and E$_{\text{cutoff}}$ is the cutoff energy of the electron population. If on the  one hand the TeV emission detected from the NW rim of Vela Jr. suggests a small value for the magnetic field, on the other hand a much higher value is needed to explain the small width of X-ray-emitting synchrotron filaments \citep{Bamba05b,Kishishita2013}. If we assume a primarily leptonic origin of the TeV emission and the electron cutoff energy of 27 TeV, which is the value inferred by \cite{2018A&A...612A...7H} for the leptonic case, we can apply Eq. \ref{eq:E_cutoff} to the values in Table \ref{tab:NW_only_CalPV}, giving an estimate on the magnetic field. We obtain that  the magnetic field can range from 2 $\mu$G in region 5 up to 16 $\mu$G in region 7. It is also possible to use the inverse argument and calculate the maximum energy reached by the electrons, in this case assuming a value for the magnetic field.

\begin{table*}[]
 \caption{Parameters derived from the NW rim fits, obtained by modeling all the \textit{XMM-Newton} spectra available for each region (PN, MOS1 and MOS2 for each observation listed in Table \ref{tab:observations}). The models tested were POWERLAW and SRCUT.
    \label{tab:NW_only_XMM}}
    \centering
\begin{tabular}{cclcc}%%
 & TBABS $\times$ POWERLAW & & TBABS $\times$ SRCUT & \\[1ex]
\hline\\[-1ex]
Region& N$\_$H (10$^{22}$ cm$^{-2}$) & \;\;\;\;$\Gamma$ &  N$\_$H (10$^{22}$ cm$^{-2}$) & Cutoff Energy (keV) \\[1ex]
\hline\\[-1ex]
 1 & 0.40$^{+0.01}_{-0.01}$ & 2.65$^{+0.01}_{-0.01}$  & 0.31$^{+0.01}_{-0.01}$& 0.28$^{+0.01}_{-0.01}$\\[1ex]
 2 & 0.27$^{+0.01}_{-0.01}$ & 2.58$^{+0.02}_{-0.02}$ &0.20$^{+0.01}_{-0.01}$ & 0.30$^{+0.02}_{-0.01}$\\[1ex]
 3 & 0.50$^{+0.01}_{-0.01}$ & 2.56$^{+0.01}_{-0.01}$ &0.42$^{+0.01}_{-0.01}$ & 0.36$^{+0.01}_{-0.01}$\\[1ex]
 4 & 0.56$^{+0.01}_{-0.01}$&  2.67$^{+0.01}_{-0.01}$ & 0.463$^{+0.004}_{-0.004}$& 0.27$^{+0.01}_{-0.01}$\\[1ex]
 5 & 0.77$^{+0.01}_{-0.01}$ & 3.11$^{+0.02}_{-0.02}$ & 0.66$^{+0.01}_{-0.01}$ & 0.11$^{+0.01}_{-0.01}$\\[1ex]
 6 & 0.57$^{+0.02}_{-0.02}$ & 2.91$^{+0.03}_{-0.03}$ &0.46$^{+0.02}_{-0.02}$ & 0.17$^{+0.01}_{-0.01}$\\[1ex]
 7 & 0.36$^{+0.01}_{-0.01}$ & 2.47$^{+0.02}_{-0.02}$ &0.28$^{+0.01}_{-0.01}$ & 0.44$^{+0.02}_{-0.02}$\\[1ex]
\hline%
\end{tabular}%
\end{table*}

\begin{table*}[]
 \caption{Parameters derived from the NW rim fits, obtained loading CalPV data. The model tested were POWERLAW and SRCUT.
    \label{tab:NW_only_CalPV}}
    \centering
\begin{tabular}{cclcc}%%
 & TBABS $\times$ POWERLAW & & TBABS $\times$ SRCUT & \\[1ex]
\hline\\[-1ex]
Region& N$\_$H (10$^{22}$ cm$^{-2}$) & \;\;\;\;$\Gamma$ &  N$\_$H (10$^{22}$ cm$^{-2}$) & Cutoff Energy (keV)\\[1ex]
\hline\\[-1ex]
1&$0.42_{-0.02}^{+0.02}$&$2.42_{-0.04}^{+0.04}$ &$0.35_{-0.02}^{+0.02}$&$0.41_{-0.05}^{+0.05}$\\[1ex]
2&$0.26_{-0.02}^{+0.02}$&$2.35_{-0.04}^{+0.04}$&$0.22_{-0.01}^{+0.01}$&$0.43_{-0.05}^{+0.05}$ \\[1ex]
3&$0.63_{-0.02}^{+0.02}$&$2.43_{-0.04}^{+0.04}$ &$0.54_{-0.02}^{+0.02}$&$0.43_{-0.04}^{+0.05}$\\[1ex]
4&$0.74_{-0.02}^{+0.02}$&$2.63_{-0.04}^{+0.03}$ &$0.63_{-0.02}^{+0.02}$&$0.27_{-0.02}^{+0.02}$\\[1ex]
5&$1.28_{-0.06}^{+0.06}$&$3.6_{-0.1}^{+0.1}$ &$0.91_{-0.02}^{+0.02}$&$0.096_{-0.002}^{+0.004}$\\[1ex]
6&$0.96_{-0.09}^{+0.08}$&$3.3_{-0.2}^{+0.2}$ &$0.63_{-0.03}^{+0.04}$&$0.133_{-0.005}^{+0.011}$\\[1ex]
7&$0.36_{-0.02}^{+0.02}$&$2.22_{-0.05}^{+0.05}$ &$0.30_{-0.02}^{+0.02}$&$0.75_{-0.11}^{+0.13}$\\[1ex]
\hline%
\end{tabular}%
\end{table*}

%
% HERE
%

%
Finally, we implemented in XSPEC the following model \citep{Zirakashvili2007,Zirakashvili2010},
\begin{equation}
F(E) \propto E^{-2} \left[1+0.38\left(\dfrac{E}{E_{0}}\right)^{0.5}\right]^{11/4}\exp\left(-\sqrt{\dfrac{E}{E_{0}}}\right)    
\label{eq:lcutoff}
,\end{equation}
which we called \texttt{lcutoff}. This model represents the spectrum of an electron distribution accelerated by a shock wave, limited by radiative losses \citep[see,  e.g.,][for a recent application]{Sapienza2022}. The relation provides a direct estimation of the expansion velocity V$_{f}$, independently of other factors than the parameter E$_{0}$: 
\begin{equation}
    E_{0}=\dfrac{2.2 \text{ keV}}{(1+\kappa^{1/2})^{2}\eta_{B}}\left(\dfrac{V_{f}}{3 \times 10^{3} \text{ km s}^{-1}}\right)^{2}
    \label{eq:E0}
.\end{equation}
The parameter E$_{0}$ depends on $\kappa$, the ratio of the upstream to the  downstream magnetic field, and $\eta_{B}$, which accounts for the deviations from the  Bohm diffusion regime. We assume $\kappa=1/\sqrt{11}$, which is, according to \cite{Zirakashvili2007},  realistic for young and nonthermal SNRs. Rearranging Eq. \ref{eq:E0}, we can write a formulation with an explicit dependence on $\eta_{B}$ (Table \ref{tab:lcutoff}). Since the expansion rate $\mu$ measured by \cite{2015ApJ...798...82A} is mostly valid for our regions 3 and 4, assuming as expansion velocity 2140 km/s (mean value $V_{f}\eta_{B}^{-1/2}$ for these two regions), we can derive the following approximate estimate of the distance ($D=V_{f}/\mu$):
\begin{equation}
     D \approx \frac{2140\,\mathrm{km/s}}{420\,\mathrm{mas/yr} }\eta_{B}^{1/2} \approx 1100 \,\eta_{B}^{1/2} \mathrm{pc}.
\end{equation}
Assuming that our assumptions and modeling results are valid, the expression obtained in this way directly relates the distance to Vela Jr to the Bohm factor $\eta_{B}$. In particular, it implies that distances significantly below $1\,\mathrm{kpc}$ can be considered unlikely, as they would require $\eta_{B} < 1$.

\begin{table}[]
 \caption{Best-fit parameters using model from Equation \ref{eq:lcutoff}.
\label{tab:lcutoff}}
\begin{tabular}{lll}%%
\centering
% Region&N$\_$H (10$^{22}$ cm$^{-2}$)&Shock velocity (km/s)\\[1ex]%
Region&N$\_$H (10$^{22}$ cm$^{-2}$)&$V_{f}\eta_{B}^{-1/2}$ (km/s)\\[1ex]%
\hline\\[-1ex]%
1&$0.36_{-0.02}^{+0.02}$&$2330_{-130}^{+160}$\\[1ex]%
2&$0.23_{-0.01}^{+0.01}$&$2460_{-180}^{+210}$\\[1ex]%
3&$0.56_{-0.02}^{+0.02}$&$2380_{-110}^{+130}$\\[1ex]%
4&$0.63_{-0.01}^{+0.02}$&$1900_{-70}^{+70}$\\[1ex]%
5&$1.03_{-0.04}^{+0.04}$&$1020_{-50}^{+50}$\\[1ex]%
6&$0.77_{-0.06}^{+0.07}$&$1110_{-90}^{+100}$\\[1ex]%
7&$0.33_{-0.02}^{+0.02}$&$3400_{-400}^{+500}$\\[1ex]
\hline
\end{tabular}
\end{table}%

\section{Distance estimate and the role of the Vela  SNR\label{sec:discussion_distance}}
Looking at our survey and pointed observation results, we obtain column densities ranging from $10^{20}$ and $10^{22}$ cm$^{-2}$, which is in agreement with the values presented in Section \ref{sec:intro}. Such a large variation can be addressed observing the data in other bands; for example, in Figure 6 of \cite{2018ApJ...866...76M} there seems to be some material overlapping the NW rim, along our line of sight, which could justify a measured column density of 10$^{21}$ cm$^{-2}$ or higher. In addition, the synchrotron emission detected by \cite{2017ApJ...850...71F} strongly correlates with the TeV emission observed by \cite{2007ApJ...661..236A} in the same region. This would justify the production of TeV photons via the hadronic channel, which requires a dense material to scatter the high energy protons coming from the SNR shock. There are two possible scenarios giving rise to the $\gamma$-ray emission \citep{2013ApJ...767...20L,2018A&A...612A...7H}. The first possibility is the so-called hadronic scenario, as the TeV photons would be produced in the decay of neutral pions, which are created during the violent collision between two protons. The second scenario, the leptonic model, involves ambient relativistic electrons upscattering seed photons up to TeV energies via the inverse-Compton (IC) mechanism. 

In conclusion, the presence of material close to the acceleration region would justify the signal observed in the  TeV band, assuming its origin is mainly hadronic. Nevertheless, some other parts of Vela Jr appear completely free of the covering HI emission. From all these different observations, we conclude that the column density is a parameter that is strongly variable across the remnant, probably due to the presence of a large amount of material along the line of sight.

\begin{figure}
    \centering
    \includegraphics[scale=0.55]{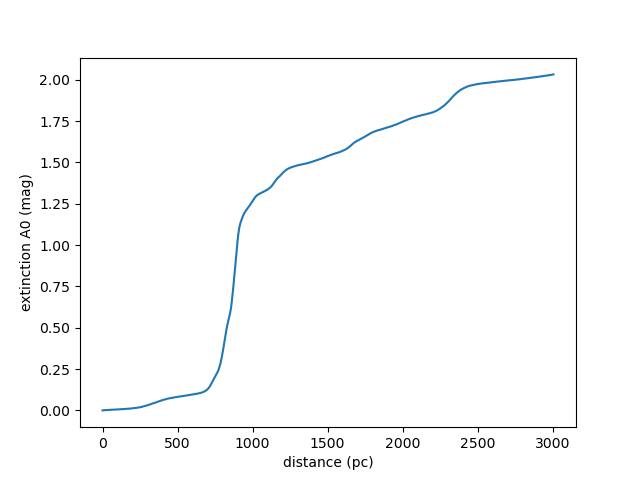}
    \caption{One-dimensional  extinction map obtained from \textit{Gaia} extinction cube and extracted using the tool \url{https://astro.acri-st.fr/gaia_dev/} in the direction of CXOU J085201.4-461753 (l=266.2$^{\circ}$, b=-1.2$^{\circ}$ in galactic coordinates)} 
    \label{fig:lallement_1D}
\end{figure}

Motivated by this evidence, we looked at the \textit{Gaia}-based extinction database given in \cite{2019A&A...625A.135L}; the observations provide parallax together with extinction measurements for each single star, giving an indication for the distance of the absorbing material. \cite{2006astro.ph..7081B} observed the CCO with \textit{XMM-Newton} and found N$_{H}=(3.8 \pm 0.3)\, 10^{21}$ cm$^{-2}$. By using the relation of \cite{1995A&A...293..889P} 
\begin{equation}
N_{H}[cm^{-2}/A_{v}]=1.79 \times 10^{21},
\label{eq:PS1995}    
\end{equation}
we obtain the expected extinction A$_{v} \sim 1.92$. Comparing this to the \textit{Gaia} data set of \cite{2019A&A...625A.135L}, this extinction corresponds to a distance of 2.4 kpc (Figure \ref{fig:lallement_1D}). Considering the radius is $\theta \sim0.95^{\circ}$ and the distance d=2.4 kpc, we can estimate the physical radius as $r=\theta \times d \sim 40$ pc. This would imply Vela Jr is a very large remnant if it turns out to be young:  this is around half the physical radius of the Vela SNR, which is at least $\sim 11$ kyr old. In conclusion, it is unlikely this is the true age of Vela Jr.

Even though the correlation between optical extinction and column density is very well established on large scales up to more than 10 kpc, there may be local deviations. Dust destruction by the shock wave of the supernova itself could be a reason \citep{2016A&A...590A..65M,2019ApJ...882..135Z} for deviations from Eq. \ref{eq:PS1995}. However, we   note that the \cite{2019A&A...625A.135L} data show a steep increase in $A_{v}$ at around 750 pc (Figure \ref{fig:lallement_1D}), which implies the distance is greater than or equal to this value unless the association of A$_{V}$ and N$_{H}$ is incorrect by a factor $\sim 10$. Given the large amount of uncertainties, this is in accordance with the lower limit set by \cite{2015ApJ...798...82A} using \textit{Chandra} expansion measurements at 500 pc. This may imply a hypothetical dust cloud at the same distance as Vela Jr, especially since massive progenitors are very likely to be found in dust-rich star-forming regions. Considering the result of \cite{2015ApJ...798...82A} and the uncertainties associated with the Eq. \ref{eq:PS1995}, the new evidence provided by the optical data of \cite{2019A&A...625A.135L} is in agreement with the findings of previous papers, leading us to conclude that a distance of Vela Jr of 750 pc or higher is very likely.

\section{Summary and discussion \label{sec:Results_Discussion}}

In Section \ref{sec:Results_Discussion} we concluded that  Vela Jr. has an overall featureless spectrum. This confirms the early indications that the NW rim of Vela Jr. is likely to be an acceleration site for high energy particles. Starting from a single power law, we tested different increasingly more complex models, specifically SRCUT and the so called \texttt{lcutoff}, to characterize in the detail each region. For the latter, we considered the electron cutoff energy derived with HESS (27 TeV for the leptonic case, considered more suitable than the hadronic scenario given the observation of X-ray synchrotron emission),finally obtaining the magnetic field can vary between 2 $\mu$G and 16 $\mu$G. Employing \texttt{lcutoff}, we manage to estimate the distance as 1100 pc, but we note how this value is a function of the Bohm factor $\eta_{B}$, which can significantly vary across the remnant \citep{Tsuji2021}. In conclusion, our findings confirm this region as probable acceleration site for particles, in accordance with the observations reporting TeV emission from the same direction. This scenario collocates Vela Jr. firmly in the class of the so called shell-type supernova remnants, with the eROSITA observation of the whole remnant confirming the results of the NW rim pointed observations.

\begin{figure}
    \centering
    \includegraphics[scale=0.6]{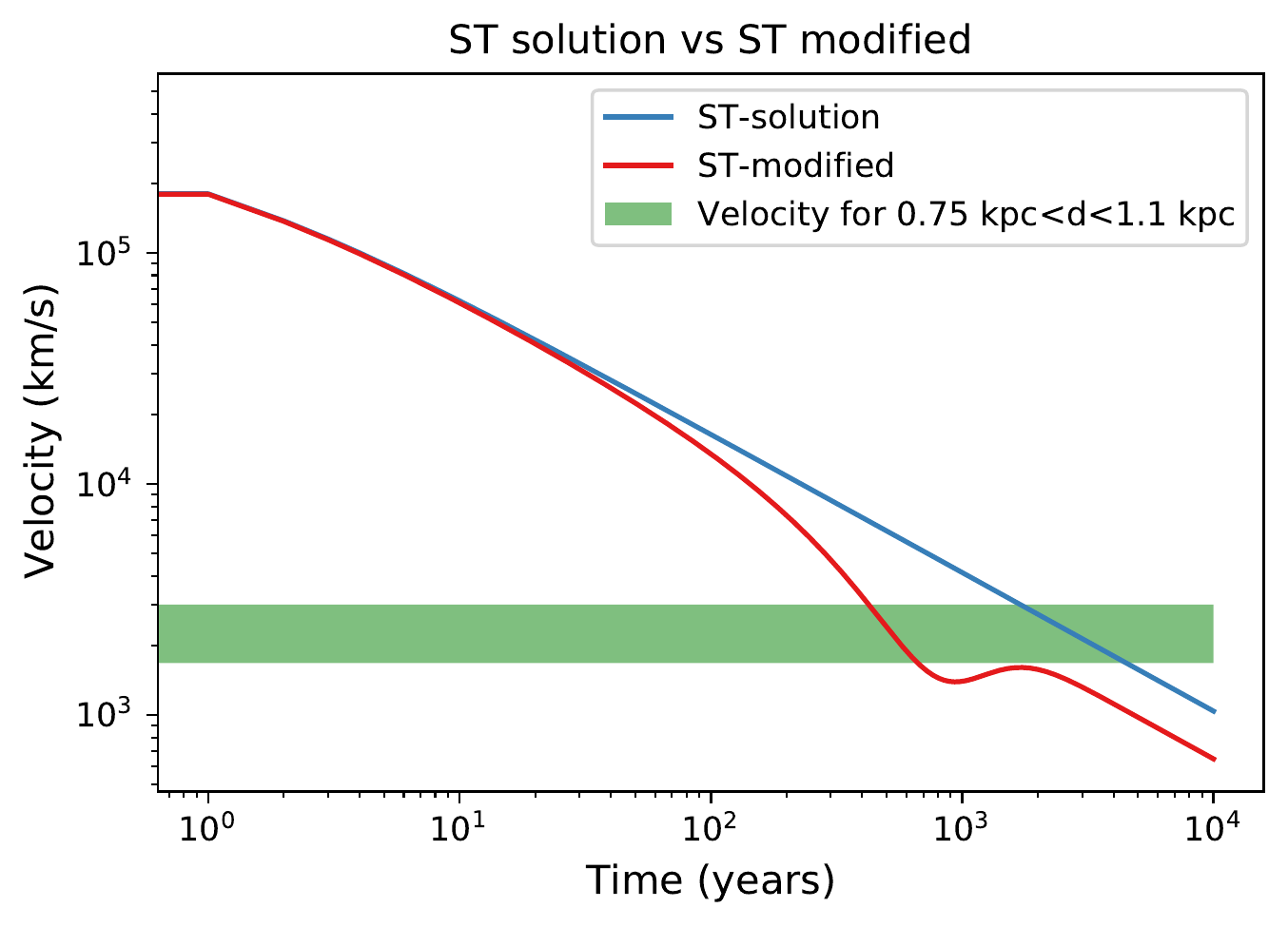}
    \caption{Expansion velocity from Sedov-Taylor (ST) equations and from the modified version of \cite{2016frap.confE..36A}. A peak appears in the latter formulation, implying a reacceleration of the front shock. The velocity range predicted for 0.75-1.1 kpc distance is also plotted. \label{fig:vST_vs_vSTmod}}
\end{figure}

Putting these findings in the broader context of TeV observations, as mentioned in Sect. \ref{sec:intro} only 8 TeV emitting sources are firmly identified with SNRs, and all of them are shell-type. This is quite a  small number if the observed number of Galactic cosmic rays is supposed to be due mainly to acceleration in shock fronts. Motivated by this observation, \cite{2016frap.confE..36A} discusses how three well-known TeV emitter SNRs, SN1006, RX 1713.7-3946 and Vela Jr may be good objects to study this discrepancy. He proposed a modified version of the Sedov-Taylor equations solution in order to account for a loss of energy in cosmic rays, instead of being dispersed in the expansion of the remnant.
Specifically, the shock front is reaccelerated after the explosion. The assumed fraction of energy transferred to cosmic rays for the modified ST version is A=0.9075 (see Figure \ref{fig:vST_vs_vSTmod}). This would place the explosion   at around 1200-1300 AD, making it compatible with the hypothesis of a much closer distance proposed by \cite{Burgess2000}. This discussion is based on the $^{44}$Ti detection reported in Section \ref{sec:intro}. However, such a small distance is ruled out by past observations (see Section \ref{sec:intro}) and our new ones (Sections \ref{sec:Spectroscopy}, \ref{sec:discussion_distance}). In conclusion, the model proposed by \cite{2016frap.confE..36A} provides an explanation for the small number of observed objects of this kind because the energy lost in particle acceleration accounts for the low surface brightness of these remnants.

However, cosmic ray acceleration as the supplier of additional momentum in the late stages of SNRs was treated with a different approach by \cite{Diesing2018}. The cosmic ray impact on the expansion of the remnant is boosted by the presence of dense ISM or even by molecular clouds. Our detection of multiple velocities and different energies in the NW rim, gives solid support to a picture with the remnant expanding in a nonuniform medium. Moreover, similar cosmic ray acceleration efficiencies were theoretically derived by \cite{Haggerty2020} and observationally tested by \cite{Giuffrida2022}. Instead of starting from Sedov-Taylor solutions, all these papers describe microscopically the transfer of energy from the shock to cosmic rays. We find that these models need a much lower transfer of energy in cosmic rays, compared to what was proposed by \cite{2016frap.confE..36A}. Additional theoretical studies connecting the Sedov-Taylor solutions with a microscopic particle acceleration treatment would be very useful to ultimately have a unified picture, eventually determining the cosmic ray acceleration efficiency. 

%allowed us for the first time to probe the diffuse emission from the inner regions of Vela Jr, aiming to distinguish it from that of Vela SNR. 
%Indeed, most of the previous studies \citep[e.g][]{2005A&A...429..225I,Bamba05b,2010ApJ...721.1492P,2008ApJ...678L..35K} were observations pointed at the rim regions or the compact objects of the region  \citep{2002ApJ...580.1060K,2006astro.ph..7081B,2013A&A...551A...7A}.

Comparing the POWERLAW and VPSHOCK models, we conclude Vela Jr. has an overall featureless spectrum with region NW, S, and W presenting flatter photon index values than inner sectors 1, 2, 3, and  4. Having steeper spectra in the inner regions is an effect likely associated with synchrotron cooling, already discussed for the case of Vela Jr. by \cite{Kishishita2013}. This is  not surprising since the shock regions should be more energetic due to the interaction with the ISM, while the inner early shocked regions should progressively cool toward the center, unless a reverse shock is in action. However, we note how the uncertainties for the  inner sector measurements are large, as shown in Figure \ref{fig:eRASS1234_powerlaw_parameters}. In conclusion, we have significant indications that the spectra of Vela Jr. are mainly nonthermal.

\begin{table}[]
 \caption{Bayesian information criterion (BIC) values for power law and power law+vpshock models and relative difference.  \label{tab:BIC_statistic}}
\begin{tabular}{ccccc}
Region&power law&vpshock+power law&$\Delta$BIC\\%
\hline%
NW&1022.29&1065.01&42.72\\%
W&1064.58&1141.54&76.96\\%
S&1096.39&1154.61&58.22\\%
sector 1&1054.45&1148.55&94.1\\%
sector 2&1465.99&1192.27&-273.72\\%
sector 3&1871.91&1168.89&-703.02\\%
sector 4&1186.46&1163.59&-22.87\\%
\end{tabular}
\end{table}

%%Comments on the shape 
%Despite both the models show very similar fit statistic for sectors 1 and 4, our preferred conclusion is that these sectors show non-thermal emission, which is distinguished from the thermal one of the Vela SNR. In addition of having reasonable values for the \texttt{powerlaw} photon index in accordance with those of the other regions, this conclusion is supported also by the spectra in Figure \ref{fig:eRASS1234_\texttt{powerlaw}_whole} and Figure \ref{fig:eRASS1234_vpshock_whole}: sector 1 shape clearly mimics a \texttt{powerlaw} when a VPSHOCK model is employed. This is probably also what gives subsolar abundances. 

%It is remarkable to notice that if the analysis was limited only to the minimization of the fit statistic, as displayed in Table \ref{tab:BF_statistic}, we would not have been able to observe significant differences between the different models. For example, for sectors 1 and 4 all the models tested give very similar statistics values, confirming these regions have very degenerate X-ray emission. Conversely, the corner plots in Figure \ref{fig:corner_sector4_\texttt{powerlaw}}, derived from the MCMC run, clearly shows well constrained and physical meaningful parameters: this does not happen with the thermal model. 

%However, since \texttt{powerlaw} gives better constraints on the parameters than a vsphock, it seems reasonable to accept this as best fitting model. 

%It would not have been possible to draw such conclusion if only an estimator like $\chi^{2}$/dof (which we approximated CSTAT/dof) was employed. 

We also observe how sector 2 and sector 3 are poorly described by a power law, and only a  two-component model definitely improves the fit. The additional thermal component is probably needed due to the presence in these regions of bright thermal filaments. Further studies,   for instance optical spectroscopy of the H$\alpha$ line in order to detect two possible velocity components, might help to associate them finally with Vela SNR or Vela Jr. 

Even though the most straightforward explanation for such different spectra is an excess of emission by Vela in that region, we cannot exclude that this change in the plasma can be associated with an extra thermal component belonging to Vela Jr. However, we observe with the two-component modeling super solar abundances in the thermal component. This is similar to abundances measured in other regions of Vela SNR by Mayer et al. (in prep.), which makes us speculate that the enhanced abundances belong to this remnant and not to Vela Jr. Further studies would be needed to clarify the origin of this additional thermal component. 

However, in the Bayesian framework we can try to assess which model fits  the data better by employing the Bayesian information criterion (BIC). It is defined as 
\begin{equation}
    \mathrm{BIC} = -2\log\, L_{\rm max}+k\log \,N
,\end{equation}
where $L_{\rm max}$ is the maximum likelihood of the model, k is the number of free parameters, and N the number of data points. We assumed the median likelihood of the model represents the maximum likelihood. According to \cite{WallJenkins2012}, the model that retrieves the smallest value of the BIC is to be considered   the best to fit the data set. From Table \ref{tab:BIC_statistic}, the power law model has to be preferred for the boundary shock regions (i.e., NW, W, and S, and sector 1). On the other hand, sectors 2 and 3 clearly favor a power law with an added thermal component, from a purely statistical point of view. For sector 4 the improvement of the fit is only marginal.
Therefore, our overall conclusion is that the spectrum of Vela Jr is uniformly featureless across all the remnant, even though (especially in sectors 2 and 3) it is unclear whether the additional thermal component can be associated with Vela Jr or Vela SNR. To test this scenario, as a consistency check we ran again the VPSHOCK fit, this time linking all the abundances to the O parameter. The result is given in Table \ref{tab:BIC_vpshock} and clearly shows how leaving all the abundance parameters (C, N, O, Ne, Mg, Si, Fe) free to vary independently is needed in sectors 2 and 3, indicating a strong variation in the plasma condition. This supports our idea that the background varies in those regions, given also the evidence provided in Section \ref{sec:Voronoi}.
\begin{table}[]
 \caption{Bayesian information criterion (BIC) values for VPSHOCK (abundances linked) and VPSHOCK models.  \label{tab:BIC_vpshock}}
\begin{tabular}{ccccc}
Region&vpshock (linked)& vpshock&$\Delta$BIC\\%
\hline%
NW&1028.4&1062.7& 34.3\\%
W&1084.76&1142.73& 57.97\\%
S&1103.6&1150.2&46.6\\%
sector 1&1026.13&1076.48&50.35\\%
sector 2&1267.92&1196.66&-71.26\\%
sector 3&1532.99&1144.93&-388.06\\%
sector 4&1136.31 &1176.43&40.12\\%
\end{tabular}
\end{table}

As presented in Section \ref{sec:Spatial_Analysis}, we addressed the question of the geometric center of the remnant. Assuming the center found by fitting a circle only through the northern remnant rim represents the true explosion site, we showed that the CCO is practically still at its birthplace. Although there are two \textit{Chandra} data sets in the public archives that were taken several years apart from each other, missing X-ray bright calibration stars in the \textit{Chandra} field of view prevented us from taking detailed proper motion measurements on the \textit{Chandra} CCO data itself. \cite{2019MNRAS.486.5716M}
proposed the existence of an IR counterpart of the CCO, which they put into question due to its missing proper motion. They set an upper limit of less than 10 mas/yr which would not be in agreement if the CCO had traveled from the \textit{ROSAT} data-based geometrical remnant center which is 4 arcmin apart (cf. \citealt{1998Natur.396..141A}). However, given the new geometrical remnant center determined in this work, the missing IR candidate proper motion is in full agreement with our finding that the CCO is practically still at its birthplace, invalidating the argument of a missing proper motion of the IR counterpart.  

\begin{figure*}
    \centering
    \includegraphics[scale=0.75]{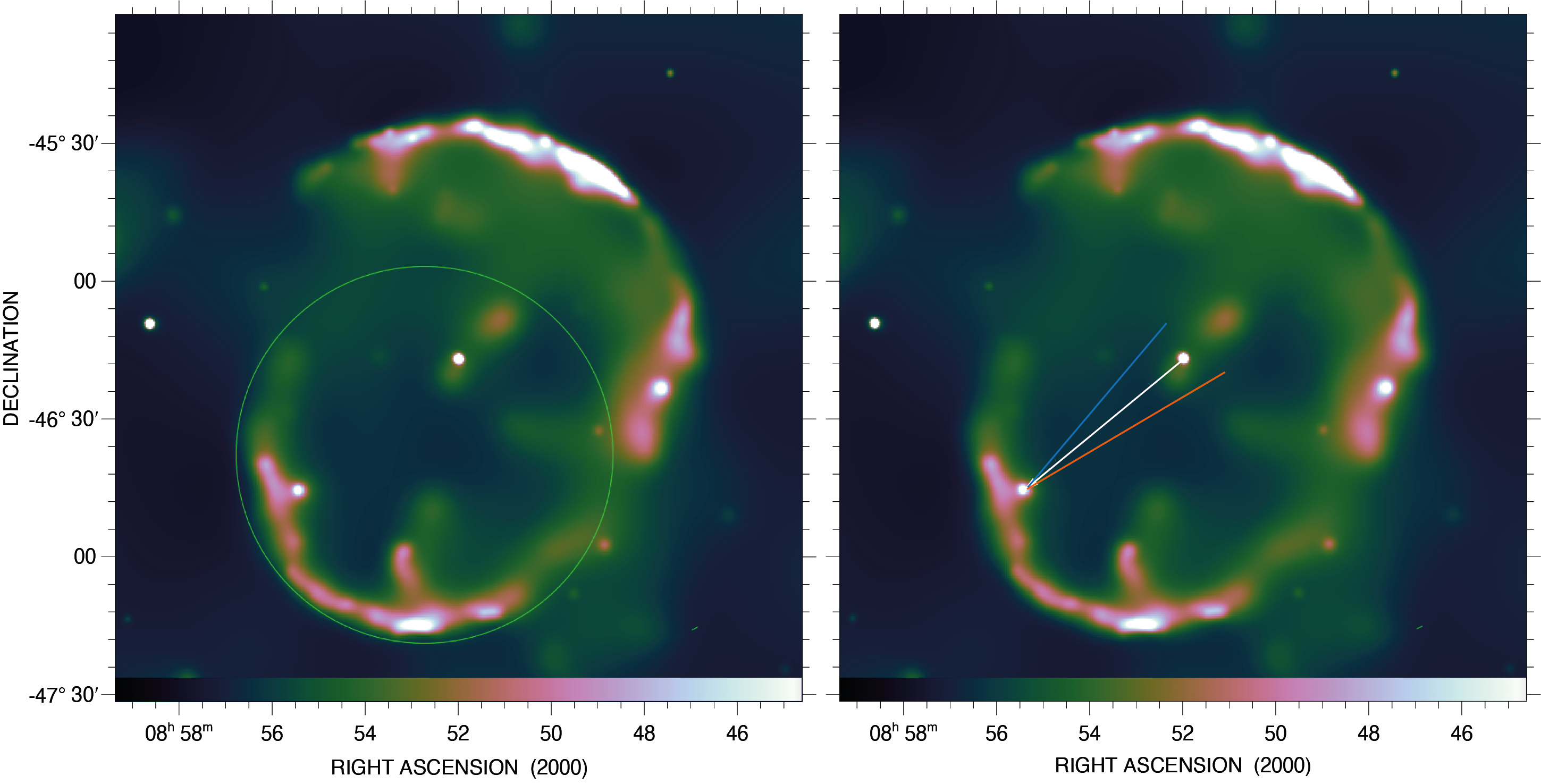}
    \caption{Vela Jr. in the $1.1-8$ keV band. The left image indicates a possible circular geometry of the southern remnant region. The circle has a radius of 41 arcmin. The right image indicates possible proper motion directions of PSR J0855-4644 as indicated from the geometry of the elongated PWN seen in the  \textit{Chandra} data (blue) by \cite{2017A&A...597A..75M} and from the geometry of the radio 
    bow shock nebular (orange) according to \cite{2018MNRAS.477L..66M}.} 
    \label{fig:VelaJr_2nd_SNR}
\end{figure*}
There are various reasons why the shape of the remnant is not exactly spherical in all places. One possible explanation for its small asymmetry is a massive progenitor, specifically a luminous blue variable (LBV). Such a progenitor would account for low density and high explosion velocities, predicted in order to have a low surface brightness remnant \citep{2016frap.confE..36A}. Remarkably, our expansion velocities measured in the NW rim (Table \ref{tab:lcutoff}) are all quite high, which fits this picture. Recent papers \citep{2021A&A...654A.167U,2021MNRAS.502..176C,2022arXiv220303369D} clearly demonstrate how   the wind emitted from a massive progenitor can also shape the circumstellar medium (CSM), and this affects the final shape of the remnant after the supernova explosion. However, a preexisting difference of density in the ISM among the different regions, especially if the remnant is large (up to 40 pc), can alternatively explain the small asymmetry.  In conclusion, the primordial differences of density in the ISM and/or the action of the wind from a massive progenitor on the CSM can be a very likely explanation for the small asymmetric shape of Vela Jr. An even simpler and probably the  most straightforward explanation would be that of an intrinsically asymmetric explosion that then leads to a deformation from a perfect spherical shape. 

On the other hand, the small asymmetry that manifests itself especially in the southern rim of Vela Jr can be described by a second circle itself, posing the question of whether Vela Jr could have been formed by two SN explosions. In this scenario, one formed the CCO and the northern part of the remnant and the second SN formed PSR J0855-4644 and the southern part of the remnant. This hypothesis justifies our initial fitting of the NW rim only. The presence of the other young and powerful rotation-powered pulsar PSR J0855-4644 within the supernova remnant boundary makes this an interesting hypothesis, which Figure~\ref{fig:VelaJr_2nd_SNR} tries to illustrate. Nonetheless, the association of PSR J0855-4644 with Vela Jr was considered unlikely by various authors in the past, mostly because of the distance mismatch between Vela Jr and the pulsar. Its radio dispersion measure-based distance was found to be 5.4 kpc, which greatly exceeds  the distance estimates for Vela Jr, regardless of all their uncertainties. 
However, \cite{2013A&A...551A...7A} concluded from a comparison of column densities a new upper limit to the distance of PSR J0855-4644, which could be as close as 900 pc, bringing the pulsar distance in a range that seems to be in agreement with the distances discussed above for Vela Jr. 
The question arises of whether  it would  be possible for a pulsar formed at the center of this putative southern circle to travel to its location in the remnant shell in a time span that is in agreement with the possible age of Vela Jr. The angular distance between the center of the southern circle shown in Figure~\ref{fig:VelaJr_2nd_SNR} and the position of PSR J0855-4644 is on the order of 0.5 deg. Assuming a distance of 1 kpc and a pulsar proper motion velocity of $V_{PSR}$ = 1000 km/s (implying a proper motion of $0.2\,\mathrm{arcsec/yr}$), it would require $\sim 8500$ yr 
to reach its location in the remnant shell. Although this assumed pulsar velocity is at the upper end of the range of possible values, considering the uncertainties in these parameters this scenario cannot be completely ruled out on the grounds of the required velocity alone. It does however seem to be in contradiction with the expansion-based age estimate of $2.4 - 5.1\,{\mathrm{kyr}}$ for Vela Jr by \citet{2015ApJ...798...82A}.

\cite{2017A&A...597A..75M} detected a compact but elongated X-ray PWN around PSR J0855-4644. Assuming that the elongated diffuse emission indicates the pulsar's proper motion direction, it would not be in agreement with a birthplace in a second southern remnant. 
Using uGMRT continuum observations of the pulsar surroundings, taken at 1.35 GHz, \cite{2018MNRAS.477L..66M} found a bow shock nebula in front of the pulsar, indicating that the pulsar's ram pressure $p_{ram}=\rho_{amb} V^2$   exceeds the ambient gas pressure, probably due to a high pulsar velocity. Although no proper motion measurement of that pulsar have been published to date, it is possible to get at least a rough estimate for its proper motion direction from the geometry of its bow shock nebula. Both estimates for the pulsar proper motion directions are indicated in Figure~\ref{fig:VelaJr_2nd_SNR}. As can be seen, they are more in agreement with a direction pointing toward  the position of the CCO rather than toward the center of a putative hypothetical second SNR which formed the southern part of Vela Jr. 
If we suppose a common origin of the CCO and PSR J0855-4644, perhaps in a binary disruption event, an important question to ask is whether the pulsar could have reached its present-day location within the age of Vela Jr. Considering the present-day distance from the CCO to the position of PSR J0855-4644 of $\sim 0.74^{\circ}$, and the measured SNR expansion rate of $(13.6\% \pm 4.2\%) \,\mathrm{kyr}^{-1}$ \citep{2015ApJ...798...82A}, even in the limit of free expansion, the required proper motion of the pulsar would be at least $(0.36\pm0.11)\mathrm{arcsec/yr}$. This however implies that the pulsar needs to have moved by around $ \sim 1.5\arcsec$ within the four years separating the two existing \textit{Chandra} observations of the object (observation IDs 13780 and 18640), which can be excluded even by a brief inspection of the two data sets. 
Thus, overall, a physical association of PSR J0855-4644 with the explosion that formed the Vela Jr SNR has to be considered quite unlikely. It is however not excluded for the progenitor stars of the pulsar and Vela Jr to have an origin in the same  star-forming region, as long as one allows for the pulsar to be significantly older than the CCO. This scenario is in agreement with the actual formation theory of CCOs \citep[see, e.g.,][]{Enoto2019}, which predicts that they are young neutron stars still embedded in the fall-back material of the supernova explosion.

\begin{acknowledgements}
 We would like to thank all the eROSITA team for the helpful discussions and suggestions provided during the realization of the paper. We thank the anonymous referee for the useful and valuable comments. MS acknowledges support from the Deutsche Forschungsgemeinschaft through the grants SA 2131/13-1, SA 2131/14-1, and SA 2131/15-1. MF acknowledges support from the Deutsche Forschungsgemeinschaft through the grant FR 1691/2-1. WB acknowledges support from the Deutsche Forschungsgemeinschaft through the grant BE 1649/11-1 and thanks Bernd Aschenbach for inspiring discussions and Luciano Nicastro for his support in data processing. FC acknowledges support from the Deutsche Forschungsgemeinschaft through the grant BE 1649/11-1 and from the International Max-Planck Research School on Astrophysics at the Ludwig-Maximilians University (IMPRS). MGFM acknowledges support from the International Max-Planck Research School on Astrophysics at the Ludwig-Maximilians University (IMPRS).
 This work is based on data from eROSITA, the soft X-ray instrument aboard \textit{SRG}, a joint Russian-German science mission supported by the Russian Space Agency (Roskosmos), in the interests of the Russian Academy of Sciences represented by its Space Research Institute (IKI), and the Deutsches Zentrum für Luft- und Raumfahrt (DLR). The \textit{SRG} spacecraft was built by Lavochkin Association (NPOL) and its subcontractors, and is operated by NPOL with support from the Max Planck Institute for Extraterrestrial Physics (MPE). The development and
construction of the eROSITA X-ray instrument was led by MPE, with contributions from the Dr. Karl Remeis Observatory Bamberg \& ECAP (FAU Erlangen-Nuernberg), the University of Hamburg Observatory, the Leibniz Institute for Astrophysics Potsdam (AIP) and the Institute for Astronomy and Astrophysics of the University of Tübingen, with the support of DLR and the Max Planck Society. The Argelander Institute for Astronomy of the University of Bonn and the
Ludwig Maximilians Universität Munich also participated in the science preparation for eROSITA. The eROSITA data shown here were processed using the
eSASS software system developed by the German eROSITA consortium. This work makes use of the Astropy Python package\footnote{\url{https://www.astropy.org/}} \citep{2013A&A...558A..33A,2018AJ....156..123A}. A particular mention goes to the in-development coordinated package of Astropy for region handling called Regions\footnote{\url{https://github.com/astropy/regions}}. We acknowledge also the use of Python packages Matplotlib \citep{Hunter:2007}, Scipy \citep{2020SciPy-NMeth}, PyLaTex\footnote{\url{https://github.com/JelteF/PyLaTeX/}} and NumPy \citep{harris2020array}. 
\end{acknowledgements}

\bibliography{bibliography}
\bibliographystyle{aa}

\appendix

\section{Background modeling and fitting procedure in eRASS:4 data \label{app:eRASS}}
Here we describe the background modeling carried out for the eRASS:4 data. In Figure \ref{fig:VelaJr_regions} we highlight background and source regions in blue and red, respectively. We extracted the background from an elliptical region around Vela Jr, representing mainly emission from Vela SNR. The spectrum extracted from this region is called \emph{X-ray background} to distinguish it from the \emph{instrumental background}. The latter is the contribution given by the sum of the internal noise and the signal created by incoming particles. More specifically, the camera is continuously hit from any direction by cosmic rays creating an X-ray continuum, produced mainly via the Bremsstrahlung effect. The components described so far are included in the instrumental background, which is not multiplied by the ARF. The components employed for the modeling were the following:
\begin{enumerate}
\item \emph{Source model}: TBABS $\times$ CONSTANT $\times$ MODEL
\item \emph{X-ray background model}: TBABS $\times$ CONSTANT $\times$  (VPSHOCK+POWERLAW)
\item \emph{instrumental model}: CONSTANT $\times$ FWC\_MODEL
\end{enumerate}
  The X-ray background is modeled using a VPSHOCK model plus a power law. We fixed the photon index of the power law, representing the contribution of the unresolved extragalactic sources (mostly AGN), to $\Gamma =1.46$. This is the standard value given in \cite{2004A&A...419..837D} and from the same paper we fixed also the normalization. This assumption was made considering that the background due to extragalactic sources should be almost spatially constant and not particularly dominant in this region, especially considering the amount of absorbing material along
the line of sight.

  The instrumental background was modeled employing the Filter Wheel Closed model (FWC) developed in \cite{Yeung2023}. This model is a combination of power laws and Gaussian lines, adapted to the processing version c020. This model was multiplied by a CONSTANT initially set to the BACKSCAL keyword (since this component is modeled without ARF). The CONSTANT parameter was left free to vary during the fit.

  Once the best fit of the background was achieved, we loaded it as background model for the source and fitted them together. The source model was multiplied by a CONSTANT set to REGAREA\footnote{See the discussion on the eROSITA Early Data Release (EDR) page for further details: \url{https://erosita.mpe.mpg.de/edr/DataAnalysis/srctool_doc.html}} keyword expressed in sr, which represents the geometric area of the region, so it is kept frozen. Therefore, the model is normalized by the area of the source.

  Without fitting the data again, we launched an \texttt{emcee} run leaving  all the  parameters of the source free to vary, while only column density (nH), temperature (kT), ionization timescale (Tau), and normalization (norm) were left free in the X-ray background. This choice was made to allow the background to slightly vary from region to region, in order to match the high variability of Vela SNR emission. For the instrumental background, only the constant in front of the FWC model is free to vary, while all the other parameters were frozen.

\section{Background modeling and fitting procedure of the XMM-Newton data \label{app:XMM}}
\begin{figure}
    \centering
    \includegraphics[scale=0.3]{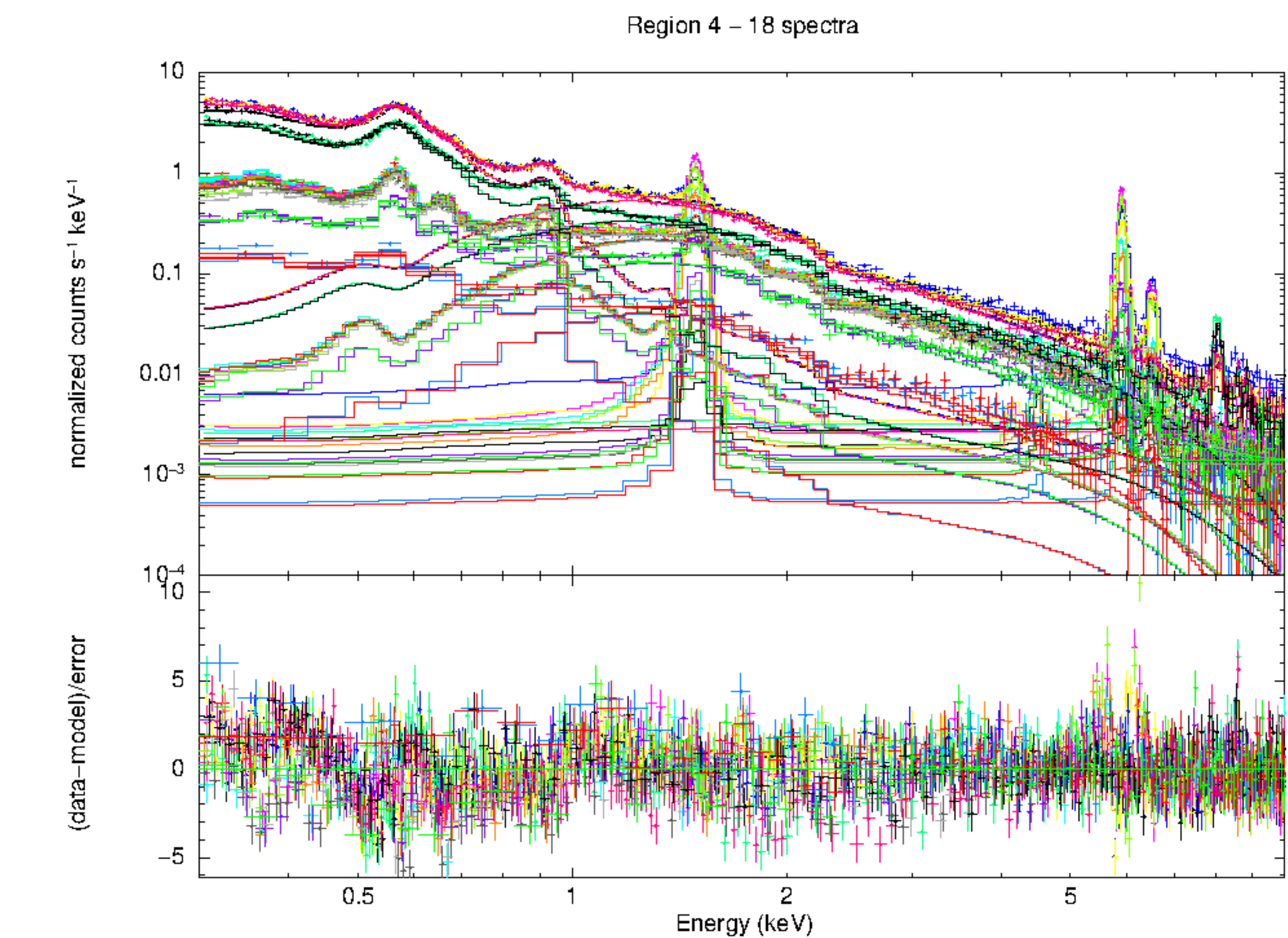}
    \caption{\textit{XMM-Newton} spectra (PN, MOS1, MOS2 for each observation, for a total of 18) from region 4 loaded and modeled. Bright emission lines are present.}
    \label{fig:18spectra_XMM}
\end{figure}

In this Appendix we   describe in detail the background modeling of \textit{XMM-Newton} data. As shown in Table \ref{tab:observations}, most of the observations are in calibration mode, which means bright emission lines are artificially created in the spectrum (Figure \ref{fig:18spectra_XMM}).
In the following we   distinguish between \emph{X-ray background} and \emph{instrumental background}. The components employed for the modeling were the following:
\begin{enumerate}
\item \emph{Source model}: TBABS $\times$ CONSTANT $\times$ MODEL
\item \emph{X-ray background model}: TBABS $\times$ CONSTANT $\times$  (VPSHOCK+VAPEC+POWERLAW)
\item \emph{Instrumental background model}: CONSTANT $\times$ (POWERLAW+GAUSSIAN lines)
\end{enumerate}
  Briefly, the instrumental background is modeled employing a superimposition of eight Gaussian lines (energies fixed  at 1.5 keV, 4.5 keV, 5.9 keV, 6.5 keV, 7.49 keV, 7.11 keV, 8.05 keV, and 8.82 keV; line width ($\sigma$) fixed to 0.001 keV) plus a power law with photon index fixed to 0.01.

  The X-ray background is mainly due to Vela SNR and is modeled as superimposition of a VAPEC, VPSHOCK, and power law models. VPSHOCK \citep{2001ApJ...548..820B} is a component accounting for thermal emission from material not in equilibrium, while VAPEC \citep{2001ApJ...556L..91S} accounts for emission from material in thermal equilibrium. Photon index and normalization were set as described in Appendix \ref{app:eRASS}, this time leaving the normalization free to vary. In addition to  the physical meaning, we looked for an empirical background model that allowed us to handle the decaying intensity and the changing shape of the spectrum.

  As the first step we loaded the background data: we ran a fit with all the norms of the model components of the instrumental and X-ray background left free to vary, with energies and width of the Gaussian lines frozen to the values listed above. This gave us the best-fit model for the background. Thus, we loaded this best-fit model of the background with the source data, freezing all the parameters except the normalization of the lines and a constant in front of the X-ray background component. This choice was made since the normalization of the lines and the continuum are variable, mainly for the action of the radioactive source, whose intensity decays with time. In the second run, we loaded only the source data that we fitted together with the background model, accounting for the variation in the instrumental background. After these steps, we froze all the parameters in the instrumental background and we launched an \texttt{emcee} run on the source model plus the global normalization of the X-ray background.

\end{document}